\documentstyle[12pt,epsfig]{article}

\advance\textheight by \topskip
\begin{document}
\tolerance=100000
\thispagestyle{empty}
\setcounter{page}{0}

\newcommand{\be}{\begin{equation}}
\newcommand{\ee}{\end{equation}}
\newcommand{\br}{\begin{eqnarray}}
\newcommand{\er}{\end{eqnarray}}
\newcommand{\ba}{\begin{array}}
\newcommand{\ea}{\end{array}}
\newcommand{\bi}{\begin{itemize}}
\newcommand{\ei}{\end{itemize}}
\newcommand{\bn}{\begin{enumerate}}
\newcommand{\en}{\end{enumerate}}
\newcommand{\bc}{\begin{center}}
\newcommand{\ec}{\end{center}}
\newcommand{\ul}{\underline}
\newcommand{\ol}{\overline}
\newcommand{\eemmbb}{$e^+e^-\rightarrow \mu^+\mu^- b\bar b$}
\newcommand{\eemmbbg}{$e^+e^-\rightarrow \mu^+\mu^- b\bar b\gamma$}
\newcommand{\eezh}{$e^+e^-\rightarrow ZH$}
\newcommand{\uub}{$ u\bar u$}
\newcommand{\ddb}{$ d\bar d$}
\newcommand{\ssb}{$ s\bar s$}
\newcommand{\ccb}{$ c\bar c$}
\newcommand{\bbb}{$ b\bar b$}
\newcommand{\ttb}{$ t\bar t$}
\newcommand{\eeb}{$ e^+ e^-$}
\newcommand{\mumub}{$ \mu^+\mu^-$}
\newcommand{\tautaub}{$ \tau^+\tau^-$}
\newcommand{\veveb}{$ \nu_e\bar\nu_e$}
\newcommand{\vmvmb}{$ \nu_\mu\bar\nu_\mu $}
\newcommand{\vtvtb}{$ \nu_\tauu\bar\nu_\tau $}
\newcommand{\lra}{\leftrightarrow}
\newcommand{\ar}{\rightarrow}
\newcommand{\sm}{${\cal {SM}}$}
\newcommand{\MH}{$M_{H}$}
\newcommand{\Dir}{\kern -6.4pt\Big{/}}
\newcommand{\Dirin}{\kern -10.4pt\Big{/}\kern 4.4pt}
\newcommand{\DDir}{\kern -7.6pt\Big{/}}
\newcommand{\DGir}{\kern -6.0pt\Big{/}}
\def\Ord{\buildrel{\scriptscriptstyle <}\over{\scriptscriptstyle\sim}}
\def\OOrd{\buildrel{\scriptscriptstyle >}\over{\scriptscriptstyle\sim}}
\def\pl #1 #2 #3 {{\it Phys.~Lett.} {\bf#1} (#2) #3}
\def\np #1 #2 #3 {{\it Nucl.~Phys.} {\bf#1} (#2) #3}
\def\zp #1 #2 #3 {{\it Z.~Phys.} {\bf#1} (#2) #3}
\def\pr #1 #2 #3 {{\it Phys.~Rev.} {\bf#1} (#2) #3}
\def\prep #1 #2 #3 {{\it Phys.~Rep.} {\bf#1} (#2) #3}
\def\prl #1 #2 #3 {{\it Phys.~Rev.~Lett.} {\bf#1} (#2) #3}
\def\mpl #1 #2 #3 {{\it Mod.~Phys.~Lett.} {\bf#1} (#2) #3}
\def\rmp #1 #2 #3 {{\it Rev. Mod. Phys.} {\bf#1} (#2) #3}
\def\xx #1 #2 #3 {{\bf#1}, (#2) #3}
\def\preprint{{\it preprint}}

\begin{flushright}
{\large DFTT 78/95}\\ 
{\large Cavendish--HEP--95/17}\\ 
{\rm November 1995\hspace*{.5 truecm}}\\ 
{\rm Revised February 1996\hspace*{.5 truecm}}\\ 
\end{flushright}

\vspace*{\fill}

\begin{center}
{\Large \bf Standard Model Higgs boson production 
and~hard~photon~radiation
in~$e^+e^-\ar \mu^+\mu^- b\bar b\gamma $~events 
at~LEP~II and Next Linear Collider}\\[0.95cm]
{\large Stefano Moretti\footnote{E-mails: Moretti@to.infn.it; 
Moretti@hep.phy.cam.ac.uk.}}\\[0.25cm]
{\it Dipartimento di Fisica Teorica, Universit\`a di Torino,}\\
{\it and I.N.F.N., Sezione di Torino,}\\
{\it Via Pietro Giuria 1, 10125 Torino, Italy.}\\[0.25cm]
{\it Cavendish Laboratory, 
University of Cambridge,}\\ 
{\it Madingley Road,
Cambridge, CB3 0HE, United Kingdom.}\\[0.2cm]
\end{center}
\vspace*{\fill}

\begin{abstract}
{\small
\noindent
We study at LEP II and 
Next Linear Collider energies Higgs production via the 
bremsstrahlung channel
$e^+e^-\ar ZH$, with $Z\ar \mu^+\mu^-$ and $H\ar b\bar b$,
and the corresponding irreducible background,
in presence of hard photon radiation, both from the initial and the
final state. We carry out an analysis that includes the computation of
all the relevant contributions to the complete tree--level matrix
element for $e^+e^-\ar \mu^+\mu^- b\bar b\gamma $ and makes use of the one
at leading order $e^+e^-\ar \mu^+\mu^- b\bar b$ 
interfaced with electron structure functions. We concentrate
on the case of mass degeneracy $M_H\approx M_Z$, for which next--to--leading
electromagnetic contributions can modify the content
of $b\bar b$--pairs coming from $H$ and $Z$ decays.
A brief discussion concerning the case $M_H\not\approx M_Z$ is also given.}
\end{abstract}

\vspace*{\fill}
\newpage
\subsection*{1. Introduction}

If the Standard Model (\sm) Higgs boson has a mass 
around $90$ GeV, it is degenerate with 
the neutral electroweak (EW) boson $Z$ ($M_Z\approx91.175$ GeV).
Although the model itself does not show any preference for this value,
such a condition could well occur in nature.
At this mass the Higgs particle predominantly decays into
$b\bar b$--pairs, and this is also a possible signature of the $Z$--vector.
Therefore, whenever the two bosons are contemporaneously produced,
their peaks overlap in the invariant mass spectrum of 
the decay products and cannot be separated, not even for optimistic
detector performances\footnote{In fact, 
the Higgs width $\Gamma_H$, for $M_H=90$ 
GeV, is $\approx2.4$ MeV, whereas experimental resolutions are generally 
comparable or larger than the $Z$ width, $\Gamma_Z=2.5$ GeV.}.
However, a closer inspection of the candidate $b\bar b$--sample (in a
window around $M_H\approx M_Z$)
should reveal a $b$--quark content larger than the one arising from $Z$--decays
only. This would then allow one to recognise the presence of the Higgs boson.
To this end, the fact that the Higgs Branching Ratio (BR) into $b$--quarks
is $\approx90\%$ (for $M_H\approx 90$ GeV), whereas $Z\ar b\bar b$
constitutes only 22\% of the possible hadronic $Z$--decays, 
is very helpful\footnote{An 
experimental method of discriminating between jets that originate from 
$b$--quarks and those produced by light quarks and gluons, namely $b$--tagging,
has been successfully exploited in the past few years, especially 
at LEP I  and the Tevatron. 
The method requires one to tag hadronic events with secondary 
vertices.}.

A Higgs boson with $M_H\approx M_Z$ was beyond the 
discovery potential of LEP I. Unless the Tevatron is able to
detected it in the short term, the first machine 
where such a particle  can be searched for 
is LEP II, provided that its Centre--of--Mass
(CM) energy reaches $\approx200$ GeV.
In fact, at this energy, the machine will be beyond the 
threshold of an on--shell $ZH$--pair, which is the dominant
Higgs production mechanism at LEP II. Recently, it has been 
assessed \cite{lep2w} that
the cleanest Higgs signature at LEP II is the one proceeding
through the decays $Z\ar \mu^+\mu^-$ and $H\ar b\bar b$ (the `golden channel'),
although also other final states can be 
considered ($H\ar\tau^+\tau^-$ and $Z\ar e^+e^-,\nu\bar\nu,\mbox{jets}$).
In the more distant future, high precision Higgs physics can be performed 
at the Next Linear Collider (NLC, with $\sqrt s=350-500$ GeV) \cite{ee500}. 
An $e^+e^-$ linear machine has a clear advantage with respect to hadron 
colliders. Because of the reduced
QCD background at the former, one can more easily exploit the $H\ar b\bar b$ 
decay channel, whereas at the latter
 extremely high $b$--tagging efficiency
and not--$b$--jet rejection are required in order to make Higgs detection
feasible\footnote{At the moment, it is not clear whether this 
will be possible or not, particularly 
in case of $Z$--$H$ 
mass degeneracy. For discussions about the Higgs discovery
potential of future hadronic machines, see Refs.~\cite{Di-teva,LHC}.}.

At LEP II the case $M_H\approx90$ GeV will require the highest
CM energy option, which corresponds to a second
stage of the project of the
machine, therefore quite far in the future.  
Furthermore, by assuming an yearly luminosity
of 500 pb$^{-1}$ for the CERN collider and
of 20 fb$^{-1}$ for the linear machine, one ends up with a number 
of events that is larger in the second case.
Also, whereas the gauge and Higgs bosons produced at LEP II would be 
practically
at rest, such that their decay products are generally soft and uniformly
distributed over the accessible phase space, at NLC energies these have 
a strong boost, making their tagging relatively easier.
For all these reasons we believe that  
the NLC will still play a decisive r\^ole in detecting and, particularly,
in studying the properties of a Higgs boson with a mass near $M_Z$.

We earlier mentioned that the $ZH$--channel
is the largest Higgs production mechanism at LEP II.
If $M_H\approx M_Z$, at $\sqrt s\approx350
$ GeV it is still the dominant process, whereas at $\sqrt s\approx500$
$WW$--fusion has bigger rates ($ZZ$--fusion is 
smaller than \eezh) \cite{fusionSM}. 
However, the $ZH$--mechanism has a few advantages 
with respect to the 
$WW$--one: its final state is fully constrained and it allows 
Higgs spin, parity and branching ratio determinations (see
\cite{BCDKZ,Orange3} and references therein).
Thus, the $ZH$--process is of a crucial importance
even in energy regimes where it is not the largest Higgs
production mechanism. This is the reason why we 
concentrate here on the bremsstrahlung process only.
In addition, we restrict our attention to the case 
$H\ar b\bar b$ (as the corresponding BR is much larger than that one into taus).
Concerning the $Z$--boson, we select muon decays only,
thus avoiding complications due to 
hadronic final states (when $Z$ goes into jets or taus), to
additional $t$--channel backgrounds (for $Z\ar e^+e^-$) and to 
the loss of kinematical constraints (final state neutrino).

The story of theoretical studies of the process \eezh\
dates back to Refs.~\cite{Valery,Ellis,Lee}, where the
Born cross section was first computed. In
Ref.~\cite{BK} the leading order (LO) 
cross section for $e^+e^- \ar f\bar f H$ was calculated analytically.
Next--to--leading order (NLO) EW
corrections were computed later 
\cite{IBA,ZZH,Full,eeHZ,GKW,nogueira}\footnote{For a review, 
see \cite{Full} (and references therein).}.
As for $M_H\approx 90$ GeV the Higgs width $\Gamma_H$ is a few MeV only,
the Higgs boson can be safely kept on--shell in the computations.
However, for phenomenological studies, the full final state, including the
Higgs decay products, has to be known. 
Therefore, signal and background processes
of the type, e.g., $e^+e^-\ar \mu^+\mu^- b\bar b$, must be considered.
Recently, tree--level semi--analytical cross sections 
for $e^+e^-\ar \mu^+\mu^-b\bar b$ events
were given \cite{primo,secondo,dima}.
Alternatively, one can exploit a Monte Carlo \cite{BCDKZ,numerical}
or a `deterministic' approach \cite{GP}.

It is the purpose of this paper to 
quantify the influence on the cross section
of hard photon emission, which can take place in the process $e^+e^-\ar 
\mu^+\mu^-b\bar b$. In fact, contrary to
LEP I, where the $Z$--resonance imposes a natural cut--off  on events
with hard photons produced by the initial state, 
at LEP II and NLC such a suppression does not act any longer and,
in addition, as the beam energy is larger, the probability
that the incoming electrons and positrons can radiate increases.
It follows that a sample of pure \eemmbb\ events, without $\gamma$--radiation,
does not exist in practise and one has to deal inevitably with 
electromagnetic ({\tt e.m.}) emission.
We also notice how all particles in the final state are 
{\sl charged} and therefore can also radiate hard photons. 

The inclusion of higher order {\tt e.m.}
effects is especially 
important in the case $M_H\approx M_Z$, since radiative corrections 
modify the relative content of $H$-- and $Z$--decays in the candidate 
$b\bar b$--sample. This is true whether one adopts 
missing mass techniques by computing 
the invariant mass recoiling against the tagged $\mu^+\mu^-$--pair,
or one attempts the reconstruction of the $b\bar b$--signal
directly from the $b$--jets \cite{GHS}. In the first case,
photons enter in the $M_{\rm{recoil}}$ spectrum and also
spoil the reconstruction of the $Z$--boson via the tagged muons, since
these latter can emit {\tt e.m.} radiation after the $Z$--decay. 
In the second case, as $\gamma$--radiation
can take place also off $b$--quark lines in the Higgs decay process,
such effects are equally present
(although with a reduced probability since
$|Q_b|=1/3<|Q_{e,\mu}|=1$ and because only half of the diagrams occurring in
case of radiation from leptons are involved here).
To avoid this, one could
try to include photons by looking at
the $M_{b\bar b\gamma}$ invariant mass. But such a procedure
would also be distorted by unwanted $\gamma$--radiation produced by
$e^+e^-$ and $\mu^+\mu^-$ pairs. Moreover, the knowledge of the absolute size
of the corrections for real photon emission
could help if one discriminates 
between a \sm\ Higgs boson 
and a possible neutral scalar from Supersymmetric models,
and/or in testing the Yukawa coupling of the Higgs particle.

In general, the hard {\tt e.m.} radiation affects the kinematics
of the $\mu^+\mu^- b\bar b$ events in various ways.
First, the presence of a hard 
photon modifies the two--body decay kinematics 
of the process $e^+e^-\ar ZH\ar
\mu^+\mu^- b\bar b$ \cite{NLCmio,split1}, 
as photon emission cannot be unambiguously assigned to
one or the other of its three possible
sources: $e^+e^-$, $\mu^+\mu^-$ and $b\bar b$.
Second, the radiated photon will affect
a cut around $M_Z$ on the $\mu^+\mu^-$--pair,  which
allows one to get rid of a large part of the backgrounds 
in the non--radiative case.
In fact, many of the configurations coming
from the $ZH$ signal do have muons whose invariant mass does not peak at
the $Z$--pole. In this respect, as 
the radiation from the $e$--lines is concentrated to a large extent along the
beam direction 
and that from
$b$--quark lines is suppressed by the charge $Q_b$, as a 
consequence, the $\mu$--lines are the main source of detected photons. 
Finally, the number of background diagrams
leading to $\mu^+\mu^- b\bar b \gamma$ final states in $e^+e^-$ initiated
processes
is very large (168 Feynman graphs at tree--level, neglecting 62
diagrams which include a suppressed $H\mu\mu$ Yukawa coupling), compared
to the number of those involving the Higgs resonance
(6 graphs at tree--level)\footnote{At
leading order in $e^+e^-\ar \mu^+\mu^- b\bar b$ events 
one has 24 and 1 diagrams, respectively, again excluding 9 diagrams
with $H\mu\mu$ couplings.}. 

In carrying out our analysis, we will make some simplifications, which
should not modify the conclusions we will get in the end.
We will perform only a tree--level calculation of the
processes $e^+e^-\ar \mu^+\mu^-b\bar b$ and
$e^+e^-\ar \mu^+\mu^-b\bar b\gamma$ (with the photon emitted
both from the initial and the final state). We will however `dress' the
former with the Initial State Radiation (ISR) \cite{ISR}, implementing this 
latter by means of a convolution with electron structure functions 
\cite{structure}
(in particular, we use the expressions given in 
Ref.~\cite{Nicro})\footnote{We neglect
to consider Linac energy spread and beamsstrahlung, as 
they are negligible compared to the ISR, at least for the `narrow'
D-D and TESLA collider designs \cite{ISR}.
These effects would however be straightforward to insert.}.
This is done in order to be able to sum up
the rates corresponding to the two above processes, 
through the order ${\cal O}(\alpha_{em}^5)$. 
In fact, the ISR formulae, as given in Ref.~\cite{Nicro}, include
the exact photon corrections to the $e^+e^-$ annihilation subprocess
up to the second order in the {\tt e.m.} coupling constant (in particular,
they also embody hard photon emission).  
For consistency, we implemented only ${\cal O}(\alpha_{em})$
ISR terms (the inclusion of the ${\cal O}(\alpha_{em}^2)$ pieces does not
change significantly the size of such corrections). 
By computing the total cross section of the leading ($2\ar4$) reaction  
supplemented by the ISR and that one of the radiative
($2\ar5$) process, we are then able to separate at the order $\alpha_{em}^5$
the contribution due
to hard photon emission from the one due to collinear and soft 
{\tt e.m.} radiation (see below
the cut in transverse momentum)
and to virtual corrections. 
In particular, we combine 
the two approaches by subtracting the radiative ($2\ar5$)
Matrix Element (ME) from the leading log ${\cal O}(\alpha_{em})$ part
of the ISR,
to avoid double counting (see also Ref.~\cite{count1,count2}).
In the following, the label `LO' will identify rates obtained from
the $2\ar4$ process dressed with ISR and with 
the hard emission subtracted,
whereas `NLO' will refer to rates obtained from the $2\ar5$ reaction
involving hard {\tt e.m.} radiation only.
This subtraction is performed between, on the one hand, the graphs in Fig.~1 
and, on the other hand, those in Figs.~2--4 in which 
photons are connected to electron and positron 
lines: i.e., 1 \& 2 in Fig.~2, 1--2, 7--9 \& 14 in Fig.~3, 
1--2 \& 8--9 in Fig.~4. 
Since in neutral current processes ISR and Final State Radiation (FSR) are 
separately gauge invariant, this procedure is legitimate here.

The above procedure is adopted for the case of the dominant,
universal, factorisable and process independent contribution
to the complete set of ISR corrections to four--fermion 
production, containing all mass singularities $L=\ln(s/m_e^2)$ and
expressed via the ISR `radiator',  as known from $s$--channel $e^+e^-$ 
annihilations\footnote{Naively, ISR universal effects tend to 
reduce the effective beam energy, thus modifying
both the normalisation and the shape  
of the differential distributions
which are of interest in Higgs searches.}.
In addition to these, 
there are also non--universal, non--factorisable and process
dependent ISR corrections, which arise
in association with $t$-- and $u$--channel 
electron exchanges (neutrino, in case of charged currents). 
The full set of formulae needed to incorporate the complete ISR
in {\tt CC3, NC2 and NC8} \cite{classification} 
$e^+e^-\ar4$ fermion processes has been recently given, in
Ref.~\cite{ISRcomplete}. Whereas Higgs signals (graph d in Fig.~1)
and non resonant background contributions (graphs b and c in Fig.~1)
proceed via $s$--channel annihilations, the remaining background
contributions (graphs a in Fig.~1) 
do through $t$-- and $u$--electron channels, involving
$ZZ$--, $Z\gamma$-- and $\gamma\gamma$--production
(i.e., {\tt NC8}--type diagrams). Thus, 
we could well apply non--universal ISR corrections to 
these latter contributions. However, it has been shown in 
Ref.~\cite{ISRcomplete} that they are generally smaller by
an order of magnitude with respect to the universal ones.
Furthermore,  one has to consider that we are concerned
with Higgs searches  for $M_H\approx M_Z$, such that  in the end 
(after the Higgs
selection procedure) the dominant background is $ZZ$--production
(i.e., {\tt NC2}--type graphs), for which the non--universal ISR
corrections are even more suppressed. For these reason then, we
neglect here such effects.

Compared to the universal ISR, QED corrections related 
to the final state (FSR) 
are much smaller (and comparable to the non--universal ISR
corrections \cite{count1,ISRcomplete,FSR,benna}). 
Typical suppressions of the order ${\cal O}(\Gamma/M)$
are expected for the interferences between ISR and FSR in 
resonant boson pair production as well as in inter--bosonic 
final state interferences \cite{interf}. We do not expect
these effects to be relevant for Higgs production either.
Thus, because of
their small size, we do not include them in our analysis.

We also ignore genuine weak corrections, as these have been proved to
be well under control (at least for the on--shell
process \eezh) \cite{genuine}.

We further stress that, on the one hand,
a calculation including even part of the above effects
would be extremely CPU--time consuming, 
because of the large number of Feynman diagrams involved in the
processes studied, which have up to seven external particles and multiple
Breit--Wigner resonances in different regions of phase space. On the
other hand, a full ${\cal O}(\alpha_{em}^5)$ calculation is well
beyond the intentions of this paper. 

A further simplification we have adopted is to avoid 
computing some of the interference terms of the two processes 
\eemmbb($\gamma$),
which either vanish identically or are extremely small
compared to the squared terms of the MEs.
Finally, we do not discuss the QCD background,
$e^+e^-\ar Z+n~{\mbox{jets}}$ (with $n\ge2$), as this is largely suppressed 
if all jets are well separated (as we assume throughout this paper)
and they 
have to reproduce the Higgs mass, $M_{\mathrm{jets}}\approx M_H$.

Finite width effects for the $Z$ and $H$ 
are included\footnote{We are not concerned about
possible gauge invariance violations due to bremsstrahlung off unstable
particles, as there is no cancellation here which could amplify 
such effects.}. 
The masses of both $b$--quarks and
muons are non--zero, apart from the case of the $H\mu\mu$ coupling, 
in which $m_\mu$
has been set equal to zero, thus eliminating the corresponding 
diagrams from the MEs.
No effort has been made to simulate experimental effects, in 
particular in detecting $b$--quarks, apart from assigning plausible
values for the
$b$--tagging efficiency $\epsilon_b$ and the misidentification
probability  ${\epsilon_c}'$ of $c$ as a $b$.
We cut on the transverse
momentum (with respect to the beam direction)
of each particle in the final state, $p_T^{\rm {all}}>1$
GeV, which, on the one hand, is reasonably compatible
with detector requirements and, on the other hand, protects our radiative 
MEs from both divergences and numerical instabilities. Hence, this value
of transverse
momentum also defines the phase space region of soft and collinear photons
(i.e., $p_T^\gamma<1$)  
emitted from the incoming $e^+e^-$--lines where 
we make use of the ISR (as described above).

In the next Section we give an account
of the numerical calculations; in Section 3
we present and discuss our results and Section 4 gives 
our conclusions.

\subsection*{2. Calculation} 

For $M_H\approx M_Z$, both at LEP II and at the
NLC the cleanest Higgs signature in the intermediate mass range will be
via the process
\be\label{eemmbb}
e^+ e^- \ar \mu^+\mu^- b\bar b.
\ee
The corresponding Feynman diagrams for background (graphs a, b and c)
and signal (graph d) contributions are depicted in 
Fig.~1\footnote{Here and in the following we adopt 
the labelling: $e^+$ (1), $e^-$ (2),
$\mu^+$ (3), $\mu^-$ (4), $b$ (5), $\bar b$ (6) and
$\gamma$ (7).}.
Among the background graphs, one distinguishes 
between `conversion diagrams' (Fig.~1a) and `annihilation diagrams'
(Fig.~1b,c). They are also called {\tt crab} and
{\tt deer} diagrams, respectively, further dividing these latter into
$\mu$--{\tt deer} (Fig.~1b) and $b$--{\tt deer} (Fig.~1c) diagrams 
\cite{primo,secondo,dima}.
By inserting the appropriate type of internal propagator ($\gamma$ or $Z$) and
by performing all the possible permutations of the boson lines on the 
fermion ones one can get out of the background diagrams of Figs.~1a--c a total
of 24 diagrams, 8 per each kind of graphs (a, b and c).
The signal diagram in Fig.~1d is unique.

To obtain the Feynman graphs representative of the radiative process
\be\label{eemmbbg}
e^+ e^- \ar \mu^+\mu^- b\bar b \gamma,
\ee
one has to add a real photon in all possible ways to the diagrams of Fig.~1.
This is a topologically 
trivial matter, however, the total number of graphs to be computed
is very large. The case of the signal is the simplest, as
one obtains the 6 diagrams of Fig.~2 (here the wavy internal lines
represent a $Z$).
For the case of the {\tt crabs} one gets 56 diagrams
in total, as indicated by Fig.~3 (in which an internal wavy line represents
both a $\gamma$ and a $Z$).
This is also the number that one gets from the {\tt deers} (both
for $\mu$ and $b$, see Fig.~4). 
In total, one ends up with 174 diagrams at tree--level.
As already mentioned, we did not consider 62 diagrams involving $H\mu\mu$
Yukawa couplings.

More than from the large number of graphs, complications in computing
the cross section for process (\ref{eemmbbg}) arise from the
fact that they have in general
rather different resonant structures over the accessible
phase space. For example (see Fig.~2), the signal diagrams
present Breit--Wigner peaks of the type $H\ar b\bar b$,
$H\ar b\bar b\gamma$, $Z\ar \mu\bar \mu$ and $Z\ar \mu\bar \mu\gamma$.
For the backgrounds, in addition to the two already 
mentioned $Z$--resonances, one
also finds $Z\ar b\bar b$ and $Z\ar b\bar b\gamma$ peaks, together with
diagrams which are not resonant at all (for example, the ones in Fig.~3
when all the internal wavy lines are identified with virtual
 photons)\footnote{As we will be eventually interested in studying
the distributions 
in the invariant masses $M_{b\bar b}$ and $M_{b\bar b\gamma}$,
for the time being
we do not discuss here the fact that also resonances
of the type $Z\ar \mu^+\mu^-b\bar b$ 
can appear in the case of the {\tt deers}, see, e.g., graph 1 
of Fig.~4 (in which
the wavy line attached to the incoming fermions represents a $Z$ and
the one joining $\mu$-- and $b$--lines is a photon).}.
The (numerical) integration must carefully take into account this
diversified structure.

We computed the cross sections
by using the packages
MadGraph \cite{tim} and HELAS \cite{HELAS} for the
Feynman diagram evaluation and VEGAS \cite{VEGAS} 
for the integration over the phase space. For VEGAS,
we split the total
Feynman amplitude squared
into a sum of non--gauge--invariant terms (as already done
in Refs.~\cite{split1,split2}), each of which  has
a particular resonant structure, and integrate them by using 
appropriate phase spaces which absorb the Breit--Wigner 
peaks in the integrand. In general, the change of variable
\be\label{change}
Q^2-M^2=M\Gamma\tan\theta, \longrightarrow
{{d}}Q^2=\frac{(Q^2-M^2)^2+M^2\Gamma^2}{M\Gamma}{{d}}
\theta,
\ee
where $Q$, $M$, $\Gamma$ stand for the virtuality, the mass and the width
of the resonance, gives an integrand which depends smoothly on $\theta$.
For example, in Fig.~2, it is convenient to isolate three terms, i.e.,
the amplitude squared of the graphs: i) 1 \& 2; ii) 3 \& 4; iii) 5 \& 6,
and to compute them separately by means of different phase spaces.
The interferences between the set of diagrams in i), ii) and iii) 
are the last contribution to the part of the
total ME which includes Higgs resonant graphs. 
In a similar way, one proceeds with
the background graphs in Figs.~3--4.

Such a procedure gives a substantial reduction of the 
integration errors, although some diagrams need
to be computed twice or more (e.g., diagrams 1 \& 2 in Fig.~2 are necessary
for the $|M_1+M_2|^2$ term as well as for some of the interferences). This
apparently increases the final CPU time of the run.
However, the competition between these two aspects (i.e., high precision
but multiple Feynman diagram evaluations) is such that 
the computing time needed to get a given accuracy is 
much less than that required to integrate the differential cross
section without any special care,
especially when one uses adaptive algorithms
for the multi--dimensional integrations, as we have here.

Among the different non--gauge--invariant terms one has to deal with,
the interferences between the
various sets (of signal, {\tt crabs}, $\mu$-- and $b$--{\tt deers})
as well as the ones between graphs within the same set
are the most complicated, as these generally mix up
graphs with different resonant structures\footnote{We have integrated them
by using a flat phase space, which does not map any of the possible
peaks of the interfering graphs.}.
However, this often implies that they are 
small compared to the amplitudes squared of diagrams with
identical composition of Breit--Wigner peaks, since the 
phase space region in which one or more graphs are large is different.
Therefore, a useful and time--saving procedure is to neglect 
such terms, whenever possible.

For example, concerning the leading order 
process (\ref{eemmbb}), it has been shown
in Ref.~\cite{dima} that the various Higgs--background interferences either 
vanish identically or are small, so that they can be safely neglected.
In the case of the interferences between the various sources
of background such a suppression is less visible \cite{primo}.
We have checked whether this remains true in presence
of hard photon radiation. For example, by interfering
the signal diagrams of Fig.~2 with the {\tt crabs} and {\tt deers} 
of Figs.~3 and 4 together, one obtains mixed 
terms which are negligible compared to all the 
squared contributions, 
whereas mixing the various
sets of {\tt crab} and {\tt deer} diagrams
of Figs.~3 and 4 yields interferences which are of the same order as
some of the amplitudes squared. 
This also
happens for the interferences within the sets of Figs.~2, 3 and 
4, if treated separately: that is, when one splits the modulus
squared of the signal process represented in Fig.~2 as previously described,
and those of the backgrounds in Figs.~3 and 4 by isolating the $Z$--resonances
out of the graphs with $\gamma$--propagators. 
In this respect, we found it useful to recognise within the 
{\tt crab} diagrams of Fig.~3 the components with the following resonant
structures: 
i)    $Z\ar \mu^+\mu^-(\gamma)$ and $Z\ar b\bar b(\gamma)$; 
ii)   $Z\ar \mu^+\mu^-(\gamma)$ and $\gamma\ar b\bar b(\gamma)$; 
iii)  $\gamma\ar \mu^+\mu^-(\gamma)$ and $Z\ar b\bar b(\gamma)$; 
iv)   $\gamma\ar \mu^+\mu^-(\gamma)$ and $\gamma\ar b\bar b(\gamma)$. 
In the following, we will refer
to them as $ZZ$, $Z\gamma$, $\gamma Z$ and $\gamma\gamma$ backgrounds.
A $Z$--resonance is present also in the diagrams of Fig.~4, 
however, we will refer to these collectively as $\mu$--
and $b$--{\tt deers} (although in the computations their 
resonant structure has been appropriately taken into account),
as these diagrams are very suppressed if compared to those 
in Fig.~3 and also because they do not have the two--to--two and 
two--to--three kinematics typical of
the signal (compare to Fig.~2). As already stressed, 
we recall here that
all these amplitude contributions we have been discussing 
must be summed up together in the end. However, if taken
separately, they provide  a useful way of
looking inside the process and distinguishing between the different
fundamental interactions. 
In summary, in the results we will present in the next section, 
we have systematically neglected interferences between the signal
and the backgrounds whereas we have kept all the others.

The following numerical values of the parameters have been adopted:
$M_{Z}=91.175$ GeV, $\Gamma_{Z}=2.5$ GeV,
$M_{W}=80.23$ GeV,
$\Gamma_{W}=2.2$ GeV, and for the 
Weinberg angle we have used its leptonic effective value of
$\sin^2_{\mathrm{eff}} (\theta_W)=0.2320$.
For the fermions:
$m_\mu=0.105$ GeV and $m_b=4.25$
GeV. 
The {\tt e.m.} coupling
constant $\alpha_{\rm{em}}$ has been set equal to 1/128. For
the Higgs width $\Gamma_{H}$ we have adopted the tree--level
expression corrected for the running of the quark masses in the vertices
$Hq\bar q$ (these have been
 evaluated at the scale $\mu=M_{H}$
\cite{running}). Therefore, in order to be consistent, we have used 
a running $b$--mass in the $Hb\bar b$ vertex of the production
process here considered. As representative values of the CM energy of LEP II
and NLC we have adopted 200, and 350, 500 GeV, respectively,
whereas the Higgs mass has been fixed at 90 GeV. 

\subsection*{3. Results}

In the non--radiative process \eemmbb\ 
Higgs signals should appear as narrow resonances in the invariant mass
spectrum of the Higgs decay products over the contribution due to
the background processes, which are generally flat, apart from
the region around $M_Z$, where the $Z$ peak (due to
the $ZZ$ {\tt crab} diagrams of Fig.~1a) clearly sticks out.
In order to detect Higgs signals via 
the two--$b$--two--$\mu$ channel one can adopt two different strategies
\cite{GHS}\footnote{We ignore the full reconstruction
of the reaction \eezh\ via the decays into jets and/or 
leptons of both the $H$-- and the $Z$--boson, by fitting their kinematics,  
because it would present additional complications, which 
are beyond the intentions of this study \cite{GHS}.}:
\begin{itemize}
\item the calculation of the missing mass recoiling against the reconstructed
$Z\ar \mu^+\mu^-$, by plotting the distribution in $M_{\rm{recoil}}=
\sqrt{[(p_{e^+}+p_{e^-})-(p_{\mu^+}+p_{\mu^-})]^2}$ and exploiting tagging
on the $\mu^+\mu^-$--system only;
\item direct reconstruction of the $H$--peak from the $b$--jets,
by plotting the distribution in $M_{b\bar b}=\sqrt{(p_{b}+p_{\bar b})^2}$. 
\end{itemize}
Once a hard photon is included via the process \eemmbbg, the next--to--leading
order contribution to the first spectrum coincides with the invariant mass
of the $b\bar b\gamma$--system, thus the rates up to
the order $\alpha_{\rm{em}}^5$
in the recoiling mass are the sum of the distributions $M_{b\bar b}$ at LO
and $M_{b\bar b\gamma}$ at NLO.
In the second approach, the full rates up to $\alpha_{\rm{em}}^5$ 
in the $M_{b\bar b}$
spectrum are given by the sum of the
$b\bar b$--invariant masses at leading and next--to--leading order.
We expect the possible alternative strategy we outlined in the Introduction
(i.e., to look at the `pure' $M_{b\bar b\gamma}$ spectrum)
to be less successful, because
in this case one would be able to correctly reconstruct Higgs peaks
only in the case of  diagrams 5 and 6 (of Fig.~2), which give a 
suppressed contribution to  the total \eemmbbg\ cross section.
In presenting our results, we concentrate then on the two spectra $M_{b\bar b}$
and $M_{b\bar b\gamma}$, with $M_H\approx M_Z$.

In order to suppress the backgrounds $\gamma Z$ and $\gamma\gamma$
(among the {\tt crabs}) as well as the contributions from the $\mu$--
and $b$--{\tt deers}, we require that the invariant mass
of the muon pair reproduces a $Z$--boson, by applying the cut, e.g., 
$\Delta {M_Z}\equiv |M_{\mu^+\mu^-}-M_Z|<10$ GeV \cite{BCDKZ,GHS}.
In this way, we mainly select the $ZZ$ contribution from 
the  {\tt crabs}, since the $Z\gamma$ one
will be largely suppressed in the end by a cut around $M_H$.
In this way, one is able to  
compare $Z\ar b\bar b$ and $H\ar b\bar b$ decays \cite{GHS}.
Furthermore, in order to reduce the content of $Z$--decays
in the candidate $b\bar b$--sample we consider events for which
$|\cos\theta_{\mu^+\mu^-}|<0.8$ \cite{GHS} (where $\theta_{\mu^+\mu^-}$ 
is the angle  of the reconstructed $Z$--boson with respect to
the beam), since
$e^+ e^-\ar ZZ$ events are strongly peaked in the forward/backward
direction due to the $t,u$--channel exchange of electrons 
\cite{cos}\footnote{This 
is also true for the cases $Z\gamma$, $\gamma Z$ and $\gamma\gamma$.}.

Tables I--III show the cross sections
in a window of 20 GeV,
centered around $M_H=90$ GeV (i.e., we consider events with $
\Delta M_H\equiv |M_{b\bar b(\gamma)}-M_H|< 10$ GeV), for the signals
$e^+e^-\ar ZH\ar \mu^+\mu^-b\bar b (\gamma)$
($S$, the
square of the last diagram in Fig.~1 at LO and of those in 
Fig.~2 at NLO) and the total backgrounds 
$e^+e^-\not\ar ZH\ar \mu^+\mu^-b\bar b (\gamma)$
($B$, the
square of the sum of the first three sets of graphs in Fig.~2 at LO
and of those in Figs.~3 and 4 at NLO), together with the total
significance $\Sigma$, for the integrated luminosity ${\cal L}\equiv
\int Ldt= 0.5(20)$ fb$^{-1}$ {\sl
per annum} at LEP~II(NLC), 
for $\sqrt s=200$ (Table~I), 350 (Table~II) and 500 GeV 
(Table~III). 
All the mentioned constraints have been applied: that is,
$p_T^{\mu,b}>1$ GeV, $\Delta M_Z<10$ GeV and 
$|\cos\theta_{\mu^+\mu^-}|<0.8$. 
In case of cross sections ($S$ and $B$ columns),
the label NLO 
identifies rates from the radiative process \eemmbbg\ only 
(with $p_T^\gamma>1$ GeV). In particular,
the second line represents rates obtained from the
distribution in invariant mass of the system $b\bar b\gamma$, whereas the
other refers to rates obtained from the 
$M_{b\bar b}$ spectrum. 
In case of significances ($\Sigma$ columns), 
the label NLO give rates obtained through the order
${\cal O}(\alpha_{em}^5)$ (all photons)
in case of the direct reconstruction method 
(upper line)
and of the missing mass analysis (lower line),
as obtained by adopting the
subtraction procedure described in the Introduction.
According to this treatment,
LO refers here to rates from 
the process \eemmbb\ corrected for universal ISR effects 
due to virtual and real photons (these latter with $p_T^{\gamma}<1$ GeV).
For reference, we also
give (in square brackets) cross sections and significances
when no radiation is present, 
as obtained by integrating the spectra in Fig.~5.

In computing the $\Sigma$'s (see Ref.~\cite{BCDKZ}), we 
neglect the (small) probability of misidentification of light 
quarks and gluons as $b$'s and we assume $BR(H\ar b\bar b)>>BR(H\ar c\bar c)$.
Then, the probability 
of picking one $b$ out of two is $[1-(1-\epsilon_b)^2]$, whereas
the total significance $\Sigma$ is
\be\label{signi}
\Sigma=\sqrt{\cal L}~\frac{\sigma(e^+e^-\ar ZH\ar \mu^+\mu^- b\bar b (\gamma))}
           {\sqrt{\sigma(e^+e^-\not\ar ZH\ar \mu^+\mu^- b\bar b (\gamma))}}
       P_b,
\ee
with
\be\label{Pb}
P_b=\frac{1-(1-\epsilon_b)^2}
         {\sqrt{[1-(1-\epsilon_b)^2]+\delta[1-(1-{\epsilon_c}')^2]}},
\ee
where the factor $\delta$ has been introduced in order to remind the reader 
that the EW couplings $\gamma^*,Z\ar q\bar q(\gamma)$ are generally
different, depending whether $q=c$ or $b$. However, to use
$\delta=1$ in formula (\ref{signi}) is always a good approximation for our 
analysis. In fact, on the one hand,
the background $B$ practically coincides at tree--level 
with the $ZZ$--piece and $BR(Z\ar b\bar b)\approx BR(Z\ar c\bar c)$. Whereas,
on the other hand, the NLO rates (for which 
$BR(Z\ar b\bar b\gamma)\approx 1/4~ BR(Z\ar c\bar c\gamma)$, although this does
not occur in all diagrams)
have to be added to the LO ones to produce the correct significances,
such that the effect of the different {\tt e.m.} coupling is largely
washed out in the end\footnote{The value of $\delta$ would be different 
from 1 when
using the NLO rates on their own, but $\Sigma$'s computed
in this way have no meaning 
for the present analysis.}.
We consider the following five combinations of $b$--tagging efficiency
($\epsilon_b$) and $c\ar b$ misidentification (${\epsilon_c}'$):
$\epsilon_b=1$ and 
${\epsilon_c}'=0$ (perfect tagging), and ${\epsilon_c}'=0.2\epsilon_b$
\cite{GHS}, for $\epsilon_b=0.2,0.4,0.6,0.8$. 

Before proceeding further, a few points 
concerning the background due to top events
are worth mentioning here. In fact, $t\bar t$--production
followed by the decay $t\bar t\ar b\bar b\mu^+\mu^-(\gamma)X$ 
gives the same signatures as 
the signal processes (both at leading and next--to--leading
order, with $X$ representing missing particles).
As $m_t$ should be around or greater than 175 GeV \cite{CDFtop,D0top},
top--antitop production will not take place at LEP II. On the contrary, 
at the NLC, $t\bar t$--pairs will be copiously produced: the study of the top 
properties (especially around the threshold) is, in fact, one of the main 
physics goals of this machine. In order to quantify at leading order 
the importance of
$e^+e^-\ar t\bar t$ events (we do not
consider non--top and single top--diagrams, as, for the purposes
of this study, they can be safely neglected \cite{tt})\footnote{We also
neglect QCD (and QED) Coulomb--like interactions between the
two top quarks at threshold \cite{vak}. Both because 
the large top mass implies that the typical spikes at $\sqrt s\approx2m_t$ do 
not show dramatically up 
in the excitation curve and also because their inclusion
would not change our conclusions about the importance of the 
$t\bar t$--background.},
we have run the code used in Refs.~\cite{split1,tt}, for the choice of
CM energies adopted here, with the decays $W^+\ar\mu^+\nu_\mu$
and $W^-\ar\mu^-\bar\nu_\mu$.
In order to  have a realistic  
estimate of the cross section at threshold for $t\bar t$--events 
at a $\sqrt s=350$ GeV NLC, 
we have taken $m_t=174$ GeV for the top mass (see Ref.~\cite{Bagliesi},
where differences $\sqrt s-2m_t\approx2$ GeV were also considered).  
As we are interested in $t\bar t\ar 
b\bar bW^+W^-$ events faking possible Higgs
signals in the $b\bar b$--channel,
we look at the rates in the $2\Delta M_H$ window. We also
apply the cuts in $\Delta M_Z$, $\cos\theta_{\mu^+\mu^-}$
and $p_T^{\mu,b}$.
Because of the relatively small BR of the charged vector
bosons into muons 
(in fact, $BR(W^-\ar \mu^-\bar\nu_\mu)\approx 11\%$), the rates for
$b\bar bW^+W^-$ events with $M_{b\bar b}$ 
around $M_H=90$ GeV are rather small. For the non--radiative process 
one finds about $2\times10^{-2}$ and
$9\times10^{-2}$ fb, at $\sqrt s=350$ and 500 GeV, respectively. 
In presence of real hard photon radiation\footnote{For the
process $e^+e^-\ar t\bar t\ar b\bar b \mu^+\mu^- \gamma X$ 
we have used 
a {\tt FORTRAN} code produced by MadGraph and HELAS interfaced
with routines generating the
$W^+\ar \mu^+\nu_\mu(\gamma)$ and $W^-\ar \mu^-\bar\nu_\mu(\gamma)$ 
decays (again, only the $t\bar t$--resonant graphs have been considered).}, 
rates are even more suppressed compared to the signal ones: they are
of the order $10^{-4}$ and $10^{-3}$ fb, respectively.
In addition, $t\bar t$--events 
have a quite large value of missing energy, i.e., $E_{\rm{miss}}\OOrd
35-40$ GeV (because of the neutrinos
from the $W$--decays, which escape the detectors), whereas the
final states $\mu^+\mu^-b\bar b(\gamma)$ are
in principle fully constrained: this
should eventually allow a further reduction in
the importance of top events. 
Therefore, we can neglect
them in the present analysis.

By looking at Tables I--III it is clear how the knowledge of the 
rates due to process (\ref{eemmbbg}) with a hard photon 
can be important
in successfully carrying out a $b$--tagging analysis, especially 
at NLC energies 
and if one adopts the direct reconstruction method.
In fact,
the signal and background LO rates obtained 
in the
window $\Delta M_H<10$ GeV  are generally
of the same
order as the corrections that one 
gets from the $M_{b\bar b}$ NLO spectra, whereas
the rates obtained from the $M_{b\bar b\gamma}$ distributions are negligible
compared to those from $M_{b\bar b}$ at LO.
The NLO corrections are relatively larger at $\sqrt s\OOrd350$ GeV than at
$\sqrt s\approx 200$ GeV. The overall effect
is a relative increase of the background component in the $b\bar b$ candidate
sample, although only of a few percent (at all values of $\sqrt s$).
The cross sections in Tables~I--III also
show how an analysis that considers only radiative events \eemmbbg\ with
$p_T^\gamma>1$ GeV 
and uses exclusively the spectrum in $M_{b\bar b\gamma}$ 
would be more complicated, as rates are at least
one order of magnitude smaller
than those obtained from the sum of LO and NLO cross sections,
and because the relative excess of signal events is reduced. Furthermore,
for the $M_{b\bar b\gamma}$ rates separately, significances would
be much smaller (see also Figs.~6--8).

Figs.~5a--c shows the differential distribution in $M_{b\bar b}$ for the
non--radiative process \eemmbb\footnote{That is, the $2\ar4$ process, without
any ISR effect. When {\tt e.m.} emission from the incoming $e^+e^-$--lines
is included (with $p_T^\gamma<1$ GeV), shapes and normalisations are
generally different. However, we have not plotted here the
corresponding curves, since in order to obtain
them in the region of interest (i.e., $\Delta M_H<10$ GeV)  
it is enough to renormalise those in Fig.~5,
according to the rates given in Tables
I--III.}, for the signal (shaded) 
and the total background.
The same sequence of cuts as in the Tables has been implemented
here (apart from
the restriction in the window $M_{b\bar b}\approx M_H$).
Rates are shown for $\sqrt s=200$ (Fig.~5a), 350 (Fig.~5b) 
and 500 GeV (Fig.~5c),
with $M_H=90$ GeV, plotting the histograms by bins of 2 GeV.
In Fig.~6--8 we present the spectrum in $M_{b\bar b}$ (a) and in
$M_{b\bar b\gamma}$ (b) in the case of the radiative process \eemmbbg\
(with $p_T^\gamma>1$ GeV),
for the signal (shaded), and the total background. The combination
of  $\sqrt s$ and $M_H$ is the same as in the previous three plots.

The interpretation of Figs.~5a--c is quite straightforward:
the signal clearly shows a narrow peak at $M_H=90$ GeV, whereas the
background (which includes the sum of the all contributions from
{\tt crabs} and {\tt deers} of Fig.~1) has a broader structure.
The $ZZ$ component contributes with 
the typical $Z\ar b\bar b$ peak, whereas the tail
at small values of $M_{b\bar b}$ is largely due to $Z\gamma$ 
background events. The steep fall of $M_{b\bar b}$ around 120 GeV at
$\sqrt s=200$ GeV is a kinematical effect due to the limited CM
energy available at LEP II, whereas the long tail for $M_{b\bar b}>90$ GeV
at the NLC is an effect due to $ZZ$ (and also $\gamma Z$) events.
The $\gamma\gamma$--term as well as the {\tt deers} and the various
interferences are smaller and do not bring any substantial 
feature into the total spectrum. In general, {\tt deers} are bigger
than the $\gamma\gamma$ {\tt crabs}, and $b$--{\tt deers} dominate over
$\mu$--ones, especially at higher energies, whereas the total
interference (of the background)
is at the same level as the $b$--{\tt deers}.
As already mentioned, the interference between signal and background
is negligible with respect to the terms discussed above and, for
simplicity, it has not been included into the figures. 

The presence of a hard photon ($p_T^\gamma>1$ GeV) significantly modifies 
the lowest order distributions in the $b\bar b$--invariant
mass (especially of the Higgs process),
see Figs.~6--8. By looking at the $M_{b\bar b}$ 
spectrum in the case of the signal one easily recognises the tail at small
invariant masses due to the $b\gamma$--component, 
whereas the $e\gamma$-- and $\mu\gamma$--diagrams
contribute to the resonance around 90 GeV. 
In the case of the spectrum in $M_{b\bar b\gamma}$
it is the other way round. Here, the Breit--Wigner peak comes from
the amplitude squared due to graphs 5 \& 6 in Fig.~2, whereas
diagrams 1 \& 2 and 3 \& 4 appear via the tail at values
of $M_{b\bar b\gamma}$ greater than 90 GeV. Although not plotted
in the figures, we note that the contributions from
the $\mu\gamma$ diagrams are comparable
to those from the $e\gamma$ ones at LEP II, whereas at NLC energies (both
at 350 and 500 GeV) the $e\gamma$ rates are larger. This is clearly
due to the effect of the cut in $p_T$ on the spectrum of the
photon produced in the initial state of the \eemmbbg\ process, as 
this latter is clearly harder at higher values of $\sqrt s$.
The contribution of $b\gamma$--diagrams is generally smaller with
respect to the previous ones by at least one order of magnitude, whereas
the interference between the three sets is completely negligible.
The effect of the {\tt e.m.} radiation on the background is not so promptly 
identifiable. However, for the distribution in $M_{b\bar b}$ one easily
recognises the $ZZ$ contribution via the peak at $M_Z$, which has contributions
from $\gamma Z$ too. The $Z\gamma$ and $\gamma\gamma$ crabs have a 
substantially flat spectrum, the same for the $b$--{\tt deers}, whereas
the $\mu$--{\tt deers} give a ${\cal O}(1\%)$ contribution to the $Z$--peak.
In the case of the $M_{b\bar b \gamma}$ distribution, 
the $Z$--peak has a long tail for $M_{b\bar b\gamma}\OOrd90$ GeV (such that
it practically disappears),
due to $b\bar b\gamma$ combinations in which the photon does not come from
a $Z\ar b\bar b\gamma$ decay.
The $\gamma\gamma$ {\tt crab} and $b$--{\tt deer} contributions are small and
flat.
In general, at NLO, {\tt crabs} are more than one order of magnitude
bigger than the {\tt deers}, with the $ZZ$ and $\gamma Z$ largely
dominant. Also for this process the (negligible) 
interferences between signal and background have not been 
plotted in the figures. 

A final remark is needed if one considers that it would generally be
impossible to tag photons too close to $b$--quarks. In fact, partons
give rise to jets with a finite angular size, such that if the photon
fails within this cone it will not be distinguished from the
other parts of the jets. Thus, its energy is counted as part of 
that one of the hadronic system associated to one of the $b$--quarks 
and the ${b\bar b}$ invariant mass 
is not experimentally measurable for \eemmbbg\ events.
Clearly, this frequently happens in the case of  
diagrams 5--6 in Fig.~2, 5--6 and 12--13 in Fig.~3 and 4. 
Occasionally, this also occurs for the graphs in which
the photon is emitted by lepton lines as well as by $b$--lines
in diagrams 7 \& 14 of Fig.~4, when it 
is collinear with one of the final $b$'s. Whereas in the first case such
effect could well be important, we expect it to be rather small in the
second one. The overall effect would be that radiative events of the type 
(\ref{eemmbbg}) would look like leading events (\ref{eemmbb}) and that 
the significance of the signals would be probably improved (as the
invariant mass of the hadronic system embodying the untagged photon
would now reproduce the Higgs mass, in the case of the mentioned diagrams 
of Fig.~2). However, in order to correctly predict such an effect one
would need to know the terms due to the virtual corrections
of the FSR along the $b$--lines, which are not computed here.
An alternative strategy, that we exploit, is to impose an additional 
cut that forces the photon to be well separated from both $b$--quarks,
for example $\cos\theta_{b\gamma}<0.95$ (corresponding to  
a cone with an angular size of $\approx18$ degrees). After applying
this additional requirement, we get that in case of the signal 
the NLO rates are reduced by $\approx19\%$ at LEP I, and of 
$\approx1\%$ and $\approx4\%$ at a NLC with $\sqrt s=350$ and 500 GeV,
respectively. The sum of the background 
suffers a reduction of $\approx7(4)[1]\%$ at 
$\sqrt s=200(350)[500]$ GeV. 
Therefore, this approach should not drastically modify
the significance of hard photon events, al least
where these are quantitatively important: that is, at the NLC.

Finally, by integrating the spectra in invariant mass at LO
and NLO one obtains that the lowest order rates are generally increased
by approximately $12(24)[27]\%$ at $\sqrt s=200(350)[500]$ GeV
for the signal, whereas for the background
from 18\% (at LEP II) to 42(60)\%
(at the NLC, with $\sqrt s=350(500)$ GeV). 
However, 
as we expect that the corrections to the invariant mass
spectra due to real photon emission behave quite similarly
to those in Figs.~6--8 
also in the case $M_H\not\approx M_Z$ (in the range, let us say, 60 GeV
$\Ord M_H\Ord$ 120 GeV), then, when 
the peaks of the $H$-- and of the $Z$--particle are well separated
(in the missing mass and/or in the $b\bar b$--invariant mass distribution),
the inclusion of radiative events in the sample should not change
the results that one obtains by an analysis at lowest order (no radiation). 
Therefore
when $M_H$ and $M_Z$ are not degenerate, and the difference
$\delta M_{HZ}=M_H-M_Z$ is larger than approximately four times
the width of the $Z$--boson (see, e.g., Ref.~\cite{GHS}), 
{\tt e.m.} radiative corrections can be safely neglected.
On the contrary, for $\delta M_{HZ}\Ord 4\Gamma_Z$, a full simulation 
is in principle
needed  in order to correctly establish the excess of $b\bar b$ decays
in the $\mu^+\mu^-b\bar b(\gamma)$ sample, especially at NLC energies.

\subsection*{4. Summary and conclusions}

We have studied next--to--leading order electromagnetic contributions
via hard photon radiation to the process \eemmbb, whose signature 
represents the `golden channel' to detect and study
the intermediate mass Higgs boson of the Standard Model 
at $e^+e^-$ colliders of the present (LEP II) and future (NLC) generation.
In fact, such events involve the production of the Higgs
particle via the bremsstrahlung process $e^+e^-\ar ZH$ followed
by the decays $Z\ar \mu^+\mu^-$ and $H\ar b\bar b$.
We restricted our attention to the $ZH$ process only, thus neglecting 
Higgs production via
the $WW$-- and $ZZ$--fusion mechanisms, because of the particular importance
of the first one
in investigating the properties of the $H$--scalar. We also
focused on the case of mass degeneracy $M_H\approx M_Z$.
In such conditions, a Higgs signal can be searched for by using
$b$--tagging techniques, to establish in $\mu^+\mu^-b\bar b $ 
samples a content of $b$--quarks much larger than the one arising 
from background $Z\ar b\bar b$ decays only.

Events of the type \eemmbbg\ should be included in the
phenomenological analysis, because at LEP II and NLC energies the initial
state itself produces many hard photons (contrary to LEP I, where
their emission is suppressed by the $Z$--width).  
Since both muons and $b$--quarks are charged particles, photons can also
be radiated in the final state. As it was not {\sl a priori} clear how
this radiation modified the integrated and differential rates of
signal and background events and their interplay,  it was
important to investigate its effects.

Two search strategies have been considered: a missing mass analysis of
the system recoiling against the $Z$--boson identified via the muon pair, 
and the direct reconstruction of the Higgs peak via the $b\bar b$--pair.
Whereas for the first procedure the inclusion of hard photons 
does not modify the $M_{\rm{recoil}}$ spectra that one gets from
the $2\ar 4$ process corrected for soft and collinear ISR,
we found that, in the second case, such radiative events 
give contributions comparable to
the differences expected at leading order between the $H\ar b\bar b$
and the $Z\ar b\bar b$ rates, for $\sqrt s\OOrd350$ GeV,
when a window around $M_H=90$ GeV is selected. In particular,
NLO contributions are generally larger in the case of the 
signal, although the overall effect is a slight reduction of the expected
relative excess of $b\bar b$--pairs from the Higgs decay. 
At LEP II the influence of higher order corrections
is smaller and, in first instance, negligible (in this case NLO rates
are even the same for signal and background). 

Therefore, in an analysis that uses the direct reconstruction 
method to disentangle Higgs signals at the NLC and
in the region $M_H\approx M_Z$, the next--to--leading order
rates should be taken into account, in order to predict correctly the
amount of $H$ and $Z$  decays into $b\bar b$--pairs.
In the case of a missing mass analysis, higher order 
electromagnetic contributions can be always safely neglected.

For values of $M_H$ different from $M_Z$, when the peaks of the two particles
are well separated, the inclusion of the {\tt e.m.} hard radiation 
should not change the conclusions one gets at lowest order. The only 
effect is an overall correction to the normalisations of the differential 
distributions, which should not drastically modify the significance factors
of the signal--to--background analysis. 

Before concluding, we remind the reader that our analysis did not make use of
a full calculation up to the ${\cal O}(\alpha_{em}^5)$ 
order\footnote{Complete QED corrections to four--fermion production 
are not available yet, although important partial results have been 
recently presented \cite{completo}.}. Only contributions
due to the tree--level ${\cal O}(\alpha_{em}^4)$  process \eemmbb\
and its $\sim \alpha_{em}$ corrections due to the complete universal ISR 
(both real and virtual)
and to real 
photon emission from the final state (with $p_T^{\gamma}>1$ GeV)  
have been included here.
Therefore, a systematic error related to the ignorance of the
virtual FSR effects and of the non factorisable corrections 
comes with our results. 
However, we notice that some of these corrections 
should go in the same direction
as our conclusions, that is, of the importance of ${\cal O}(\alpha_{em})$
effects on the invariant mass spectra.
For example, according to the Kinoshita--Lee--Nauenberg theorem \cite{KLN},  
one expects that logarithmic contributions due to `collinear'
emission of photons from the final 
state 
must not appear in the  expression of the cross sections,
these being canceled by the (negative) contributions due to
virtual final state photons. In the present paper, the former
are included\footnote{At least for the $\mu^+\mu^-$ 
system, if one implements the cut in 
$\cos\theta_{b\gamma}$.} 
 whereas the latter are not: therefore, the size of what
we called NLO terms {\sl over}--estimates that 
of the real $\sim\alpha_{em}$ FSR corrections in the total cross section.
Conversely, since such negative terms have the same kinematics as the 
$2\ar4$ process, they tend to reduce the LO piece. In particular,
the Higgs peaks in the $M_{b\bar b}$ distributions 
should be less pronounced in the end. Thus,   
neglecting these virtual terms corresponds to somewhat
{\sl under}--estimating the 
effect of hard photon emission on the distributions that are of interest
for Higgs searches. 

\subsection*{Acknowledgments}

We are grateful to Ezio Maina for reading the manuscript and for useful
suggestions.
This work is supported in part by the
Ministero dell' Universit\`a e della Ricerca Scientifica, 
by the UK Particle Physics and Astronomy Research Council and by the EC
Programme ``Human Capital and Mobility'', contract CHRX--CT--93--0357
(DG 12 COMA).


\subsection*{Table Captions}

\begin{description}

\item[Tab.~I  ] Cross sections in femtobarns 
in the window $|M_{b\bar b(\gamma)}-M_H|< 10$ GeV (being $M_H=90$ GeV),
after the cuts:
$p_T^{\mu,b}>1$ GeV, $|M_{\mu^+\mu^-}-M_Z|<10$ GeV and 
$|\cos\theta_{\mu^+\mu^-}|<0.8$,
for signal ($S$) and background ($B$), together with the total
significance $\Sigma$ (see page 10--11 in the text for
their definitions), for the 
integrated luminosity ${\cal L}= 0.5$ 
fb$^{-1}$. 
The label LO refers to the rates of the 
leading process (\ref{eemmbb}) corrected
for the ISR due to soft, collinear and virtual photons, whereas
the label NLO identifies those from process (\ref{eemmbbg}) with
hard photons ($p_T^\gamma>1$ GeV).
The second line for NLO events represents rates derived from the
distributions in $M_{b\bar b\gamma}$, whereas the
others refer to cross sections and significances obtained from the 
$M_{b\bar b}$ spectra. 
The CM energy is $\sqrt s=200$ GeV.

\item[Tab.~II ] Same as Tab.~I, with $\sqrt s=350$ GeV and ${\cal L}= 20$ 
fb$^{-1}$. 

\item[Tab.~III] Same as Tab.~I, with $\sqrt s=500$ GeV and ${\cal L}= 20$ 
fb$^{-1}$. 

\end{description}

\subsection*{Figure Captions}

\begin{description}

\item[Fig.~1 ] Feynman diagrams at tree--level for the process \eemmbb\
(with the labelling: $e^+$ (1), $e^-$ (2),
$\mu^+$ (3), $\mu^-$ (4), $b$ (5) and $\bar b$ (6)): 
a) {\tt crabs}; b) $\mu$--{\tt deers};
c) $b$--{\tt deers} and d) the signal $ZH$. Internal wavy lines represent
a $\gamma$ or a $Z$ in the case of the graphs a, b and c, whereas for
the signal they represent a $Z$ only. Permutations of the virtual boson
lines along the fermion ones are not shown.

\item[Fig.~2 ] Feynman diagrams at tree--level for the process \eemmbbg\
(with the labelling: $e^+$ (1), $e^-$ (2),
$\mu^+$ (3), $\mu^-$ (4), $b$ (5), $\bar b$ (6) and
$\gamma$ (7)) in the case of the signal $ZH$.
Internal wavy lines represent a $Z$.

\item[Fig.~3 ] Feynman diagrams at tree--level for the process \eemmbbg\
(with the labelling: $e^+$ (1), $e^-$ (2),
$\mu^+$ (3), $\mu^-$ (4), $b$ (5), $\bar b$ (6) and
$\gamma$ (7)) in the case of the {\tt crabs}.
Internal wavy lines represent a $\gamma$ or a $Z$.

\item[Fig.~4 ] Feynman diagrams at tree--level for the process \eemmbbg\
(with the labelling: $e^+$ (1), $e^-$ (2),
$\mu^+$ (3), $\mu^-$ (4), $b$ (5), $\bar b$ (6) and
$\gamma$ (7)) in the case of the {\tt deers}.
Labels $i[i]$ (for $i=3,4,5$ and 6) identify the $b[\mu]$--{\tt deer}
contributions. Internal wavy lines represent a $\gamma$ or a $Z$.

\item[Fig.~5 ] Distribution in the invariant mass of the $b\bar b$--pair
($M_{b\bar b}$) in non--radiative
\eemmbb\ events, for signal (shaded) and
background (see page 10 in the text for their definitions),
 for $M_H=90$ GeV, $\sqrt s=200$ GeV (a), $\sqrt s=350$ GeV (b) 
and $\sqrt s=500$ GeV (c), 
after the sequence of cuts:
$p_T^{\mu,b}>1$ GeV, $|M_{\mu^+\mu^-}-M_Z|<10$ GeV and 
$|\cos\theta_{\mu^+\mu^-}|<0.8$. 
 
\item[Fig.~6 ] Distribution in the invariant mass of the $b\bar b$--
($M_{b\bar b}$, a) and of the $b\bar b\gamma$--system ($M_{b\bar b\gamma}$, b) 
in \eemmbbg\ events, for signal (shaded) and
background (see page 10 in the text for their definitions), 
for $M_H=90$ GeV and $\sqrt s=200$ GeV,
after the sequence of cuts:
$p_T^{\rm{all}}>1$ GeV, $|M_{\mu^+\mu^-}-M_Z|<10$ GeV and 
$|\cos\theta_{\mu^+\mu^-}|<0.8$. 
 
\item[Fig.~7 ] Same as Fig.~6, with $\sqrt s=350$ GeV.

\item[Fig.~8 ] Same as Fig.~6, with $\sqrt s=500$ GeV.

\end{description}
\vfill
\newpage

\begin{table}
\begin{center}
\begin{tabular}{|c|c|c|c|c|c|c|}
\hline
\multicolumn{7}{|c|}
{\rule[0cm]{0cm}{0cm}
$e^+e^-\ar \mu^+\mu^- b\bar b(\gamma)$}
 \\ \hline  \hline
\rule[0cm]{0cm}{0cm}
$S$ (fb) & $B$ (fb) &\omit $~$ &\omit 
$~$ &\omit $~\Sigma$ &\omit $~$ & $~$ \\ \hline
\rule[0cm]{0cm}{0cm}
$~$ &
$~$ & 
$\epsilon_b=0.2$ & 
$\epsilon_b=0.4$ & 
$\epsilon_b=0.6$ & 
$\epsilon_b=0.8$ & 
$\epsilon_b=1.0$ \\ \hline\hline
\rule[0cm]{0cm}{0cm}
$10.0[14.4]$ & 
$6.17[8.92]$ & 
$1.55[1.87]$ & 
$2.01[2.45]$ & 
$2.32[2.78]$ & 
$2.44[2.92]$ & 
$2.85[3.42]$ \\ \hline
\multicolumn{7}{|c|}
{\rule[0cm]{0cm}{0cm}
LO[no radiation]} \\ \hline  \hline
\rule[0cm]{0cm}{0cm}
$1.13$ & 
$1.10$ & 
$1.59$ & 
$2.10$ & 
$2.38$ & 
$2.51$ & 
$2.92$ \\ \hline
\rule[0cm]{0cm}{0cm}
$0.94$ & 
$0.92$ & 
$1.58$ & 
$2.10$ & 
$2.37$ & 
$2.50$ & 
$2.91$ \\ \hline
\multicolumn{7}{|c|}
{\rule[0cm]{0cm}{0cm}
NLO}
\\ \hline  \hline
\multicolumn{7}{|c|}
{\rule[0cm]{0cm}{0cm}
$|M_{b\bar b(\gamma)}-M_H|<10$ GeV\quad
\quad\quad\quad\quad$p_T^{\mu,b}>1$ GeV}
\\ \hline  \hline
\multicolumn{7}{|c|}
{\rule[0cm]{0cm}{0cm}
$|M_{\mu^+\mu^-}-M_Z|<10$ GeV\quad
\quad\quad\quad\quad$|\cos\theta_{\mu^+\mu^-}|<0.8$ }
\\ \hline  \hline
\multicolumn{7}{|c|}
{\rule[0cm]{0cm}{0cm}
$\sqrt s=200$ GeV
\quad\quad\quad$M_H=90$ GeV
\quad\quad\quad${\cal L}=0.5$ fb$^{-1}$}
\\ \hline 
\multicolumn{7}{c}
{\rule{0cm}{1cm}
{\Large Tab. I}}  \\
\multicolumn{7}{c}
{\rule{0cm}{0cm}}
\end{tabular}
\end{center}
\end{table}

\vfill
\clearpage

\begin{table}
\begin{center}
\begin{tabular}{|c|c|c|c|c|c|c|}
\hline
\multicolumn{7}{|c|}
{\rule[0cm]{0cm}{0cm}
$e^+e^-\ar \mu^+\mu^- b\bar b(\gamma)$}
 \\ \hline  \hline
\rule[0cm]{0cm}{0cm}
$S$ (fb) & $B$ (fb) &\omit $~$ &\omit 
$~$ &\omit $~\Sigma$ &\omit $~$ & $~$ \\ \hline
\rule[0cm]{0cm}{0cm}
$~$ &
$~$ & 
$\epsilon_b=0.2$ & 
$\epsilon_b=0.4$ & 
$\epsilon_b=0.6$ & 
$\epsilon_b=0.8$ & 
$\epsilon_b=1.0$ \\ \hline\hline
\rule[0cm]{0cm}{0cm}
$4.04[4.79]$ & 
$1.53[1.94]$ & 
$7.94[8.35]$ & 
$10.5[11.0]$ & 
$11.9[12.5]$ & 
$12.5[13.2]$ & 
$14.6[15.4]$ \\ \hline
\multicolumn{7}{|c|}
{\rule[0cm]{0cm}{0cm}
LO[no radiation]} \\ \hline  \hline
\rule[0cm]{0cm}{0cm}
$0.95$ & 
$0.66$ & 
$8.21$ & 
$10.9$ & 
$12.3$ & 
$12.9$ & 
$15.1$ \\ \hline
\rule[0cm]{0cm}{0cm}
$0.29$ & 
$0.17$ & 
$8.06$ & 
$10.7$ & 
$12.1$ & 
$12.7$ & 
$14.8$ \\ \hline
\multicolumn{7}{|c|}
{\rule[0cm]{0cm}{0cm}
NLO}
\\ \hline  \hline
\multicolumn{7}{|c|}
{\rule[0cm]{0cm}{0cm}
$|M_{b\bar b(\gamma)}-M_H|<10$ GeV\quad
\quad\quad\quad\quad$p_T^{\mu,b}>1$ GeV}
\\ \hline  \hline
\multicolumn{7}{|c|}
{\rule[0cm]{0cm}{0cm}
$|M_{\mu^+\mu^-}-M_Z|<10$ GeV\quad
\quad\quad\quad\quad$|\cos\theta_{\mu^+\mu^-}|<0.8$ }
\\ \hline  \hline
\multicolumn{7}{|c|}
{\rule[0cm]{0cm}{0cm}
$\sqrt s=350$ GeV
\quad\quad\quad$M_H=90$ GeV
\quad\quad\quad${\cal L}=20$ fb$^{-1}$}
\\ \hline 
\multicolumn{7}{c}
{\rule{0cm}{1cm}
{\Large Tab. II}}  \\
\multicolumn{7}{c}
{\rule{0cm}{0cm}}
\end{tabular}
\end{center}
\end{table}

\vfill
\clearpage

\begin{table}
\begin{center}
\begin{tabular}{|c|c|c|c|c|c|c|}
\hline
\multicolumn{7}{|c|}
{\rule[0cm]{0cm}{0cm}
$e^+e^-\ar \mu^+\mu^- b\bar b(\gamma)$}
 \\ \hline  \hline
\rule[0cm]{0cm}{0cm}
$S$ (fb) & $B$ (fb) &\omit $~$ &\omit 
$~$ &\omit $~\Sigma$ &\omit $~$ & $~$ \\ \hline
\rule[0cm]{0cm}{0cm}
$~$ &
$~$ & 
$\epsilon_b=0.2$ & 
$\epsilon_b=0.4$ & 
$\epsilon_b=0.6$ & 
$\epsilon_b=0.8$ & 
$\epsilon_b=1.0$ \\ \hline\hline
\rule[0cm]{0cm}{0cm}
$1.89[2.12]$ & 
$0.62[0.79]$ & 
$5.80[5.78]$ & 
$7.68[7.63]$ & 
$8.70[8.65]$ & 
$9.16[9.11]$ & 
$10.7[10.6]$ \\ \hline
\multicolumn{7}{|c|}
{\rule[0cm]{0cm}{0cm}
LO[no radiation]} \\ \hline  \hline
\rule[0cm]{0cm}{0cm}
$0.51$ & 
$0.38$ & 
$5.80$ & 
$7.67$ & 
$8.68$ & 
$9.15$ & 
$10.7$ \\ \hline
\rule[0cm]{0cm}{0cm}
$0.094$ & 
$0.064$ & 
$5.81$ & 
$7.68$ & 
$8.69$ & 
$9.16$ & 
$10.7$ \\ \hline
\multicolumn{7}{|c|}
{\rule[0cm]{0cm}{0cm}
NLO}
\\ \hline  \hline
\multicolumn{7}{|c|}
{\rule[0cm]{0cm}{0cm}
$|M_{b\bar b(\gamma)}-M_H|<10$ GeV\quad
\quad\quad\quad\quad$p_T^{\mu,b}>1$ GeV}
\\ \hline  \hline
\multicolumn{7}{|c|}
{\rule[0cm]{0cm}{0cm}
$|M_{\mu^+\mu^-}-M_Z|<10$ GeV\quad
\quad\quad\quad\quad$|\cos\theta_{\mu^+\mu^-}|<0.8$ }
\\ \hline  \hline
\multicolumn{7}{|c|}
{\rule[0cm]{0cm}{0cm}
$\sqrt s=500$ GeV
\quad\quad\quad$M_H=90$ GeV
\quad\quad\quad${\cal L}=20$ fb$^{-1}$}
\\ \hline 
\multicolumn{7}{c}
{\rule{0cm}{1cm}
{\Large Tab. III}}  \\
\multicolumn{7}{c}
{\rule{0cm}{0cm}}
\end{tabular}
\end{center}

\end{table}
\vfill
\clearpage

\begin{figure}[p]
~\epsfig{file=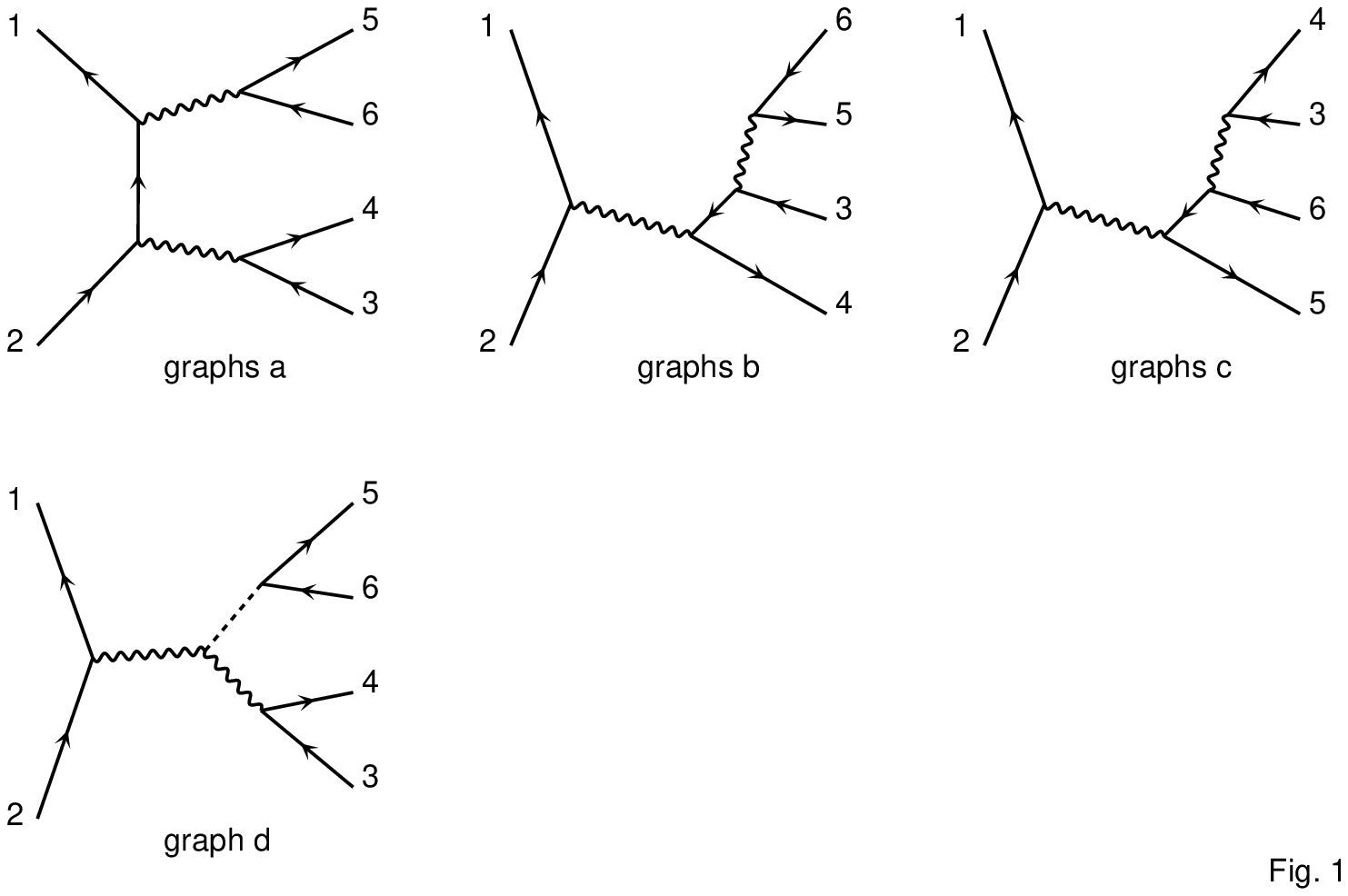,height=22cm}
\end{figure}
\stepcounter{figure}
\vfill
\clearpage

\begin{figure}[p]
~\epsfig{file=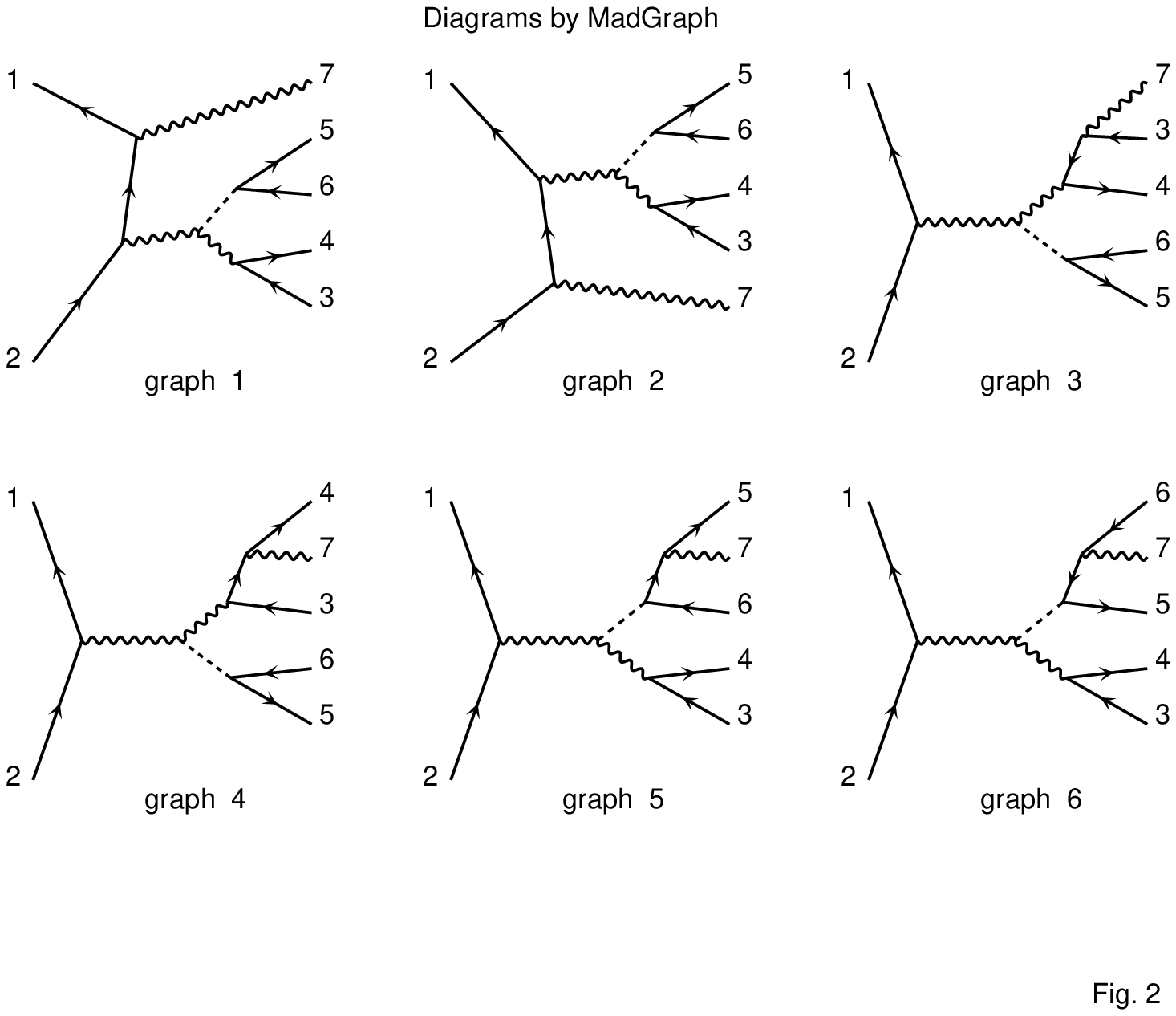,height=22cm}
\end{figure}
\stepcounter{figure}
\vfill
\clearpage

\begin{figure}[p]
~\epsfig{file=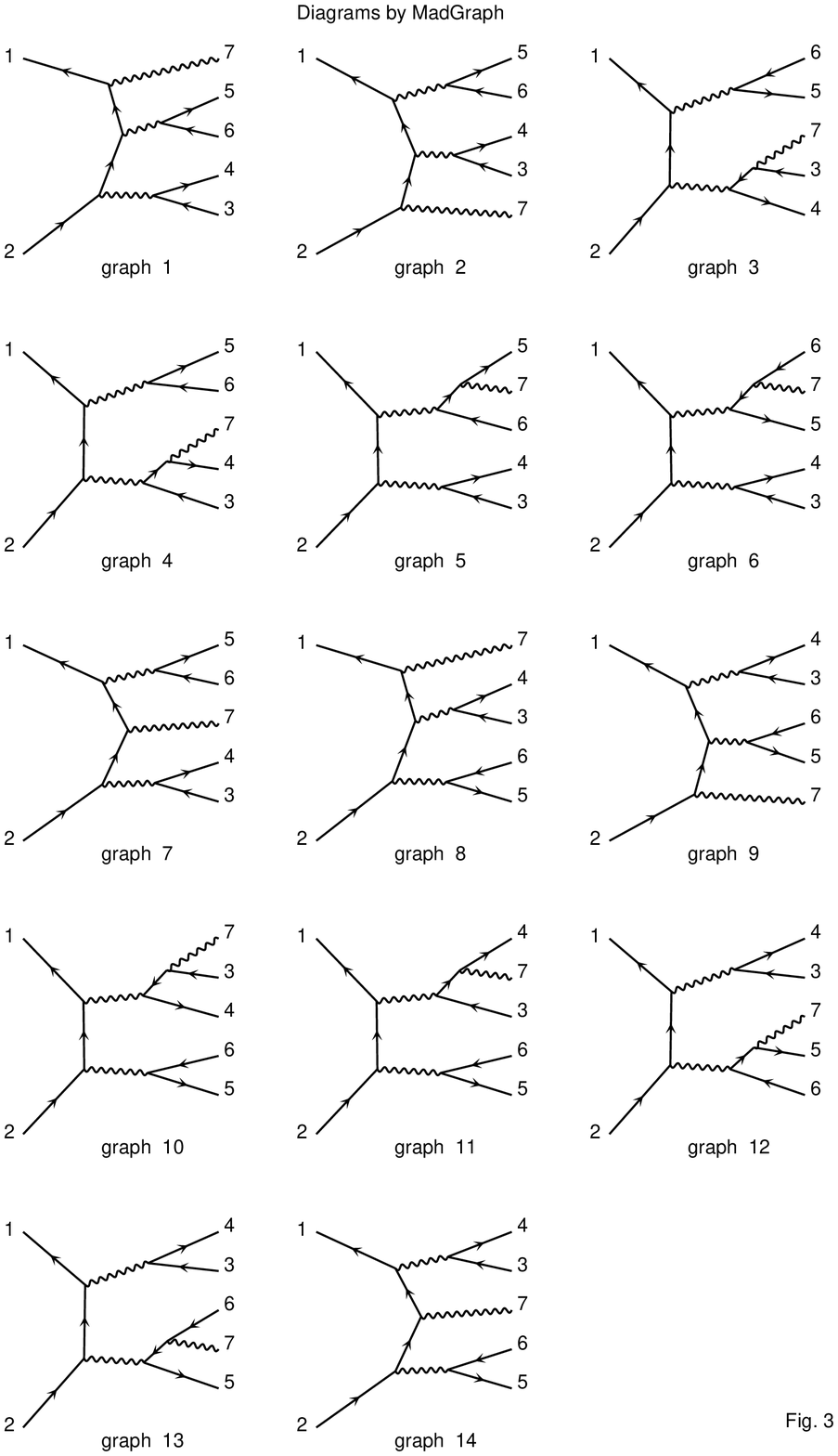,height=22cm}
\end{figure}
\stepcounter{figure}
\vfill
\clearpage

\begin{figure}[p]
~\epsfig{file=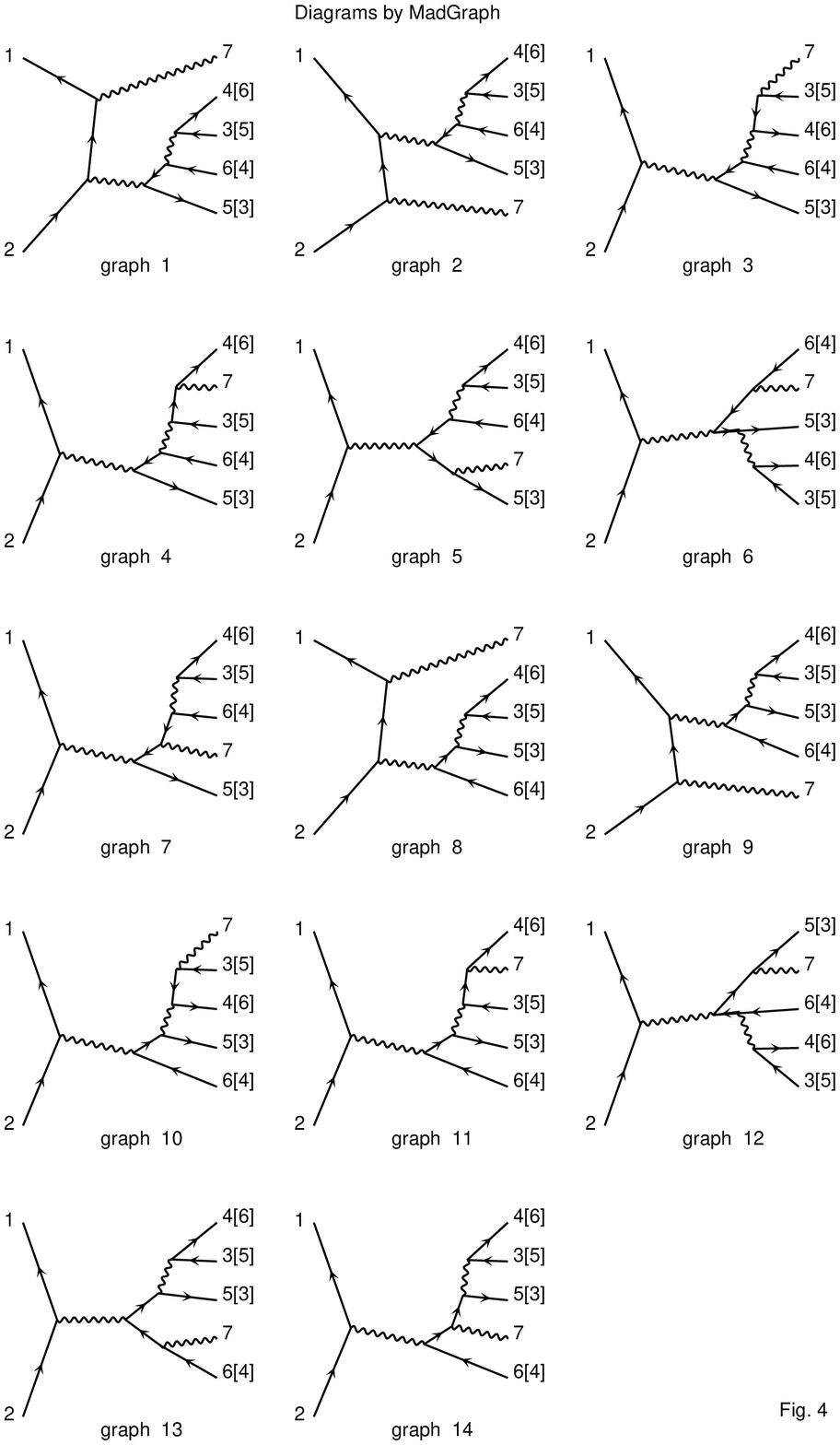,height=22cm}
\end{figure}
\stepcounter{figure}
\vfill
\clearpage

\begin{figure}[p]
~\epsfig{file=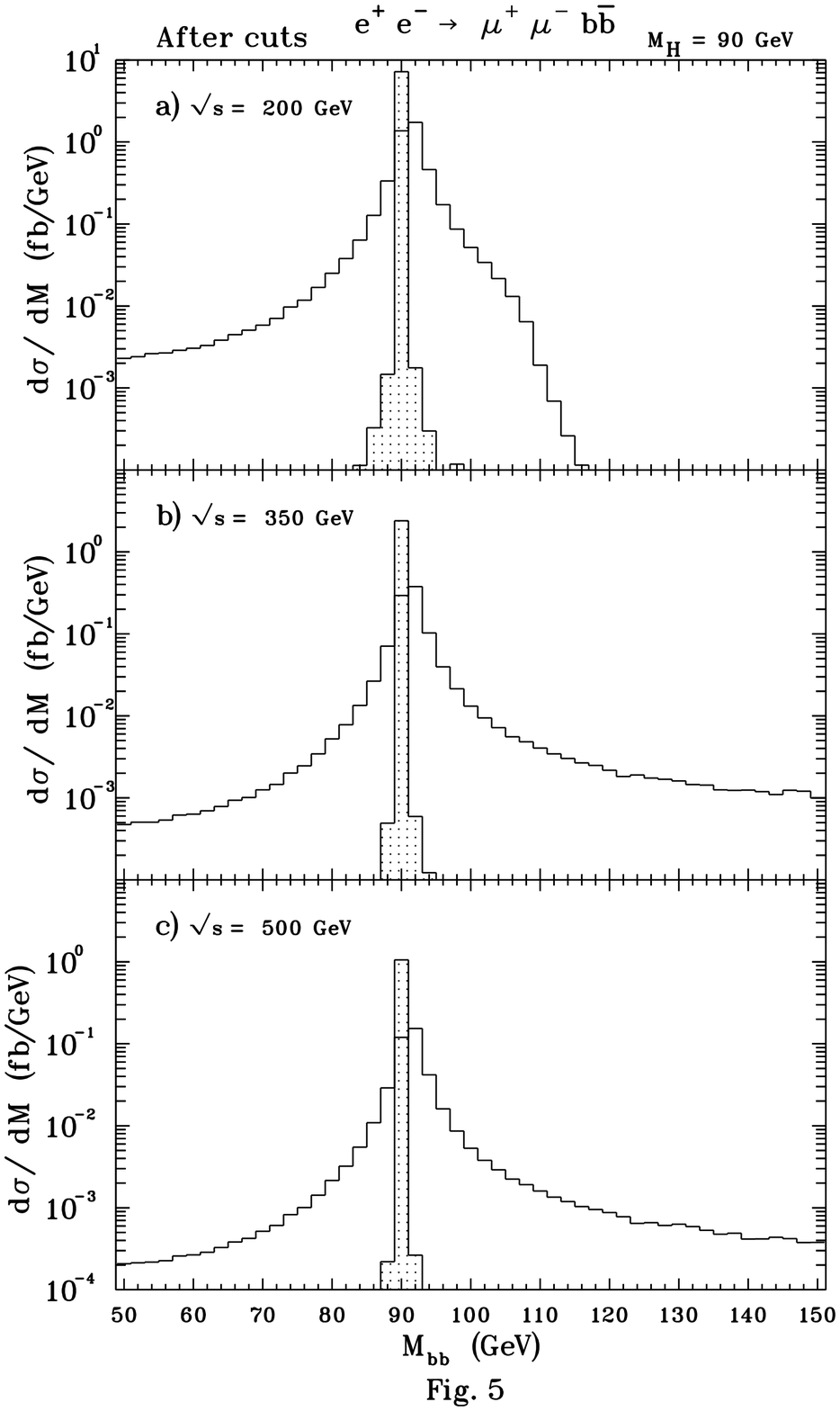,height=22cm}
\end{figure}
\stepcounter{figure}
\vfill
\clearpage

\begin{figure}[p]
~\epsfig{file=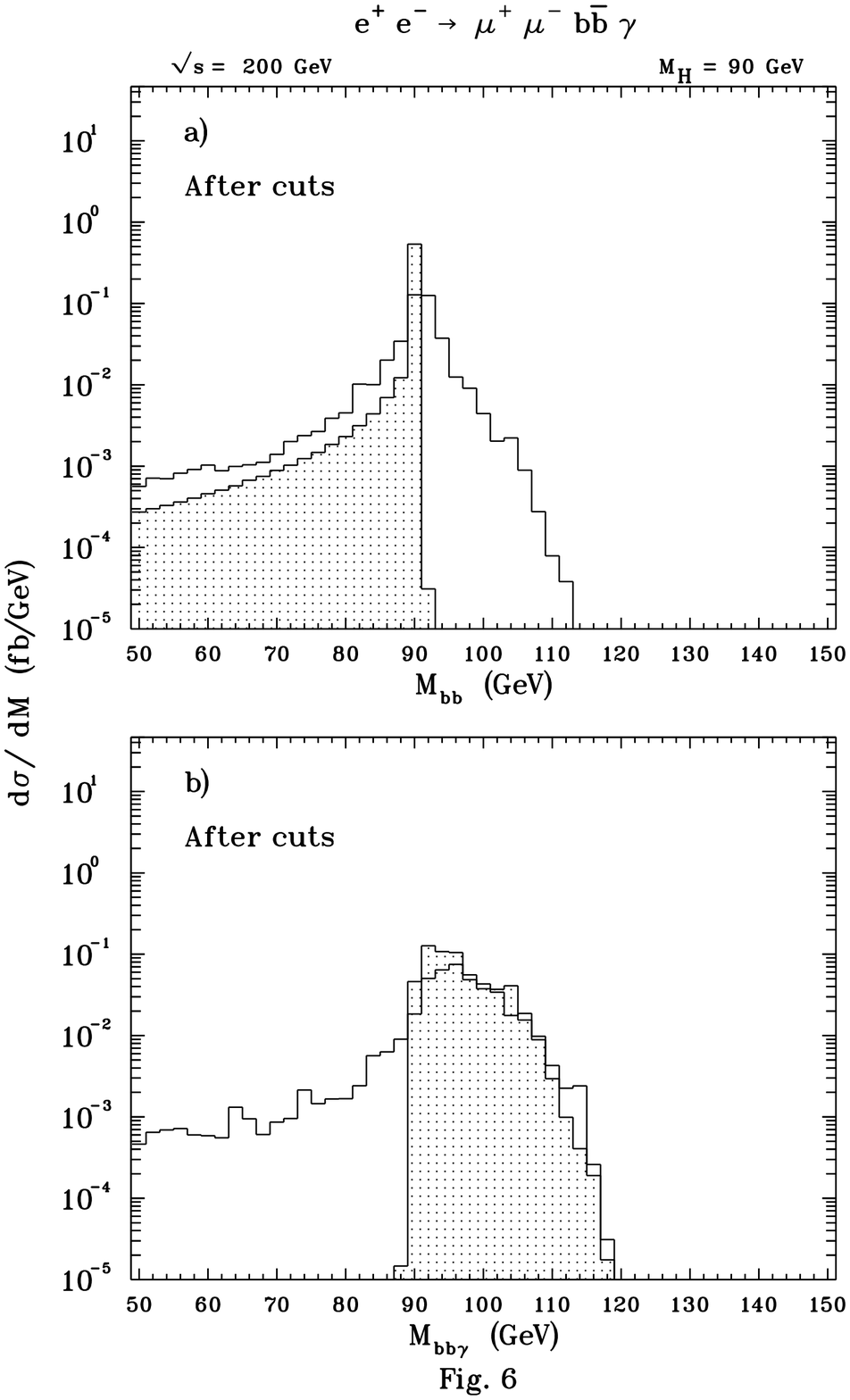,height=22cm}
\end{figure}
\stepcounter{figure}
\vfill
\clearpage

\begin{figure}[p]
~\epsfig{file=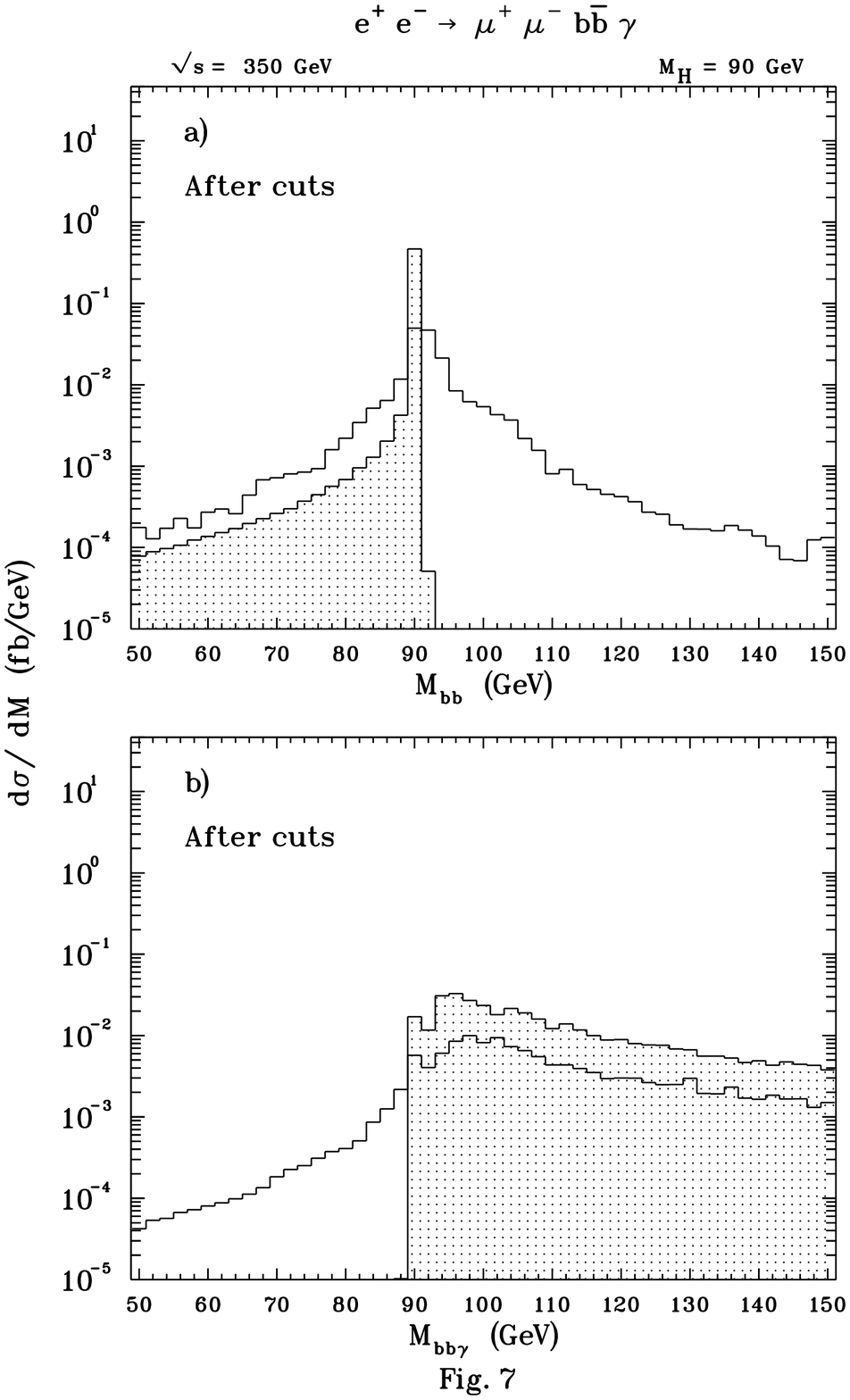,height=22cm}
\end{figure}
\stepcounter{figure}
\vfill
\clearpage

\begin{figure}[p]
~\epsfig{file=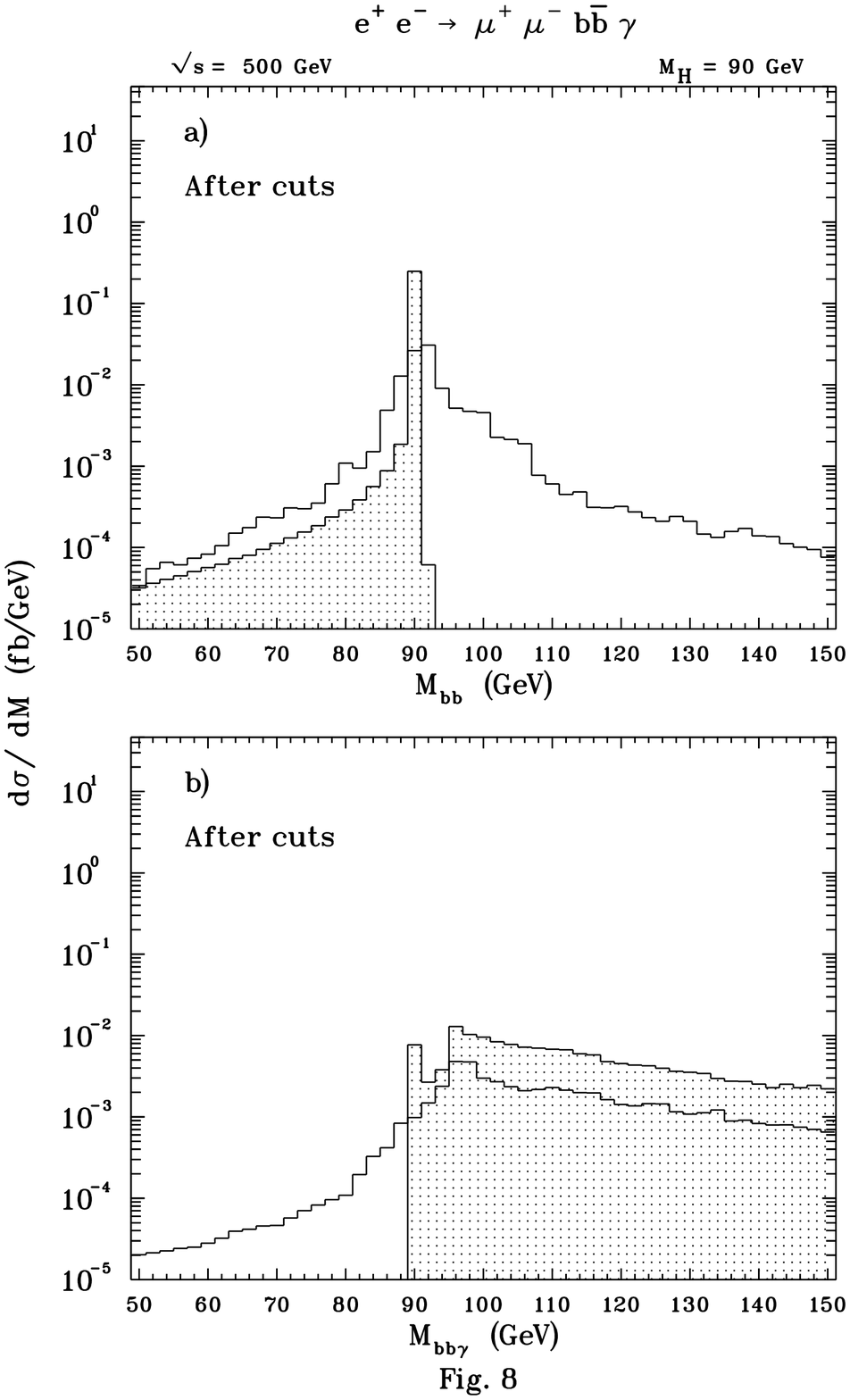,height=22cm}
\end{figure}
\stepcounter{figure}
\vfill
\clearpage

\vfill

\begin{thebibliography}{99}

\bibitem{lep2w}
G. Altarelli, T. Sj\"ostrand and F. Zwirner, eds., 
`{\it Report of the Workshop  on  Physics at LEP2}', CERN 96--01 (1996). 

\bibitem{ee500} Proceedings of the Workshop `{\it $e^+e^-$ 
Collisions at 500 GeV. The Physics Potential}',
                Munich, Annecy, Hamburg, 3--4 February 1991, ed. P.M.~Zerwas,
                DESY pub. 92--123A/B/C, August 1992.

\bibitem{Di-teva}  G.~Jackson, talk presented at the Workshop on {\it
Electroweak Symmetry Breaking at TeV-Scale Physics}, UC-Santa Barbara,
February 1994.

\bibitem{LHC} Proceedings of the `{\it
Large Hadron Collider Workshop}', Aachen, 4--9 October
1990, eds. G.~Jarlskog
and D.~Rein, Report CERN 90--10, ECFA 90--133, Geneva, 1990;\\
ATLAS Technical Proposal,
CERN/LHC/94-43 LHCC/P2 (December 1994);\\
CMS Technical Proposal, CERN/LHC/94-43 LHCC/P1 (December 1994).

\bibitem{fusionSM} D.R.T.~Jones and S.T.~Petkov, \pl B84 1979 440;\\
                   R.N.~Cahn and S.~Dawson, \pl B136 1984 196;\\
                   K.~Hikasa, \pl B164 1985 341;\\
                   G.~Altarelli, B.~Mele and F.~Pitolli, \np B287 1987 205;\\
                   B.~Kniehl, {\it preprint} DESY 91--128, 1991.

\bibitem{BCDKZ} V.~Barger, K.~Cheung, A.~Djouadi, B.A.~Kniehl and P.M.~Zerwas,
                \pr D49 1994 79.

\bibitem{Orange3}  V.~Barger, K.~Cheung, A.~Djouadi, B.A.~Kniehl,
                   R.J.N.~Phillips and
                   P.M.~Zerwas, in  Ref.~\cite{ee500}.

\bibitem{Valery} B.L.~Ioffe and V.A.~Khoze, 
                 {\it Sov. J. Part. Nucl.} {\bf 9} (1978) 50.

\bibitem{Ellis}  J.~Ellis, M.K.~Gaillard and D.V.~Nanopoulos, 
                 \np B106 1976 292.

\bibitem{Lee} B.W.~Lee, C.~Quigg and H.B.~Thacker, \pr D16 1977 1519.

\bibitem{BK} F.A.~Berends and R.~Kleiss, \np B260 1985 32.

\bibitem{IBA} M.~Consoli, W.F.L.~Hollik and F.~Jegerlehner, in 
              `{\it Z Physics at LEP 1}', ed. by G.~Altarelli, R.~Kleiss and
               C.~Verzegnassi, CERN Yellow Report No. 89-08 (1989),
               Vol.~1, page 7;\\
              W.F.L.~Hollik, {\it Fortschr. Phys.} {\bf 38} (1990) 165.

\bibitem{ZZH} Z.~Hioki, \pl B224 1989 417.

\bibitem{Full} B.A.~Kniehl, \pl B282 1992 249.

\bibitem{eeHZ} J.~Fleischer and F.~Jegerlehner, \np B216 1983 469;\\
                 B.A.~Kniehl, \zp C55 1992 605;\\
                 A.~Denner, J.~Kublbeck, R.~Mertig and M.~B\"ohm, \zp
                 C56   1992 261.
\bibitem{GKW} E.~Gross, G.~Wolf and B.A.~Kniehl, {\it Z. Phys.} 
              {\bf C63} (1994) 417; Erratum, {\it ibidem} {\bf C66}
              (1995) 321.

\bibitem{nogueira} P. Nogueira and J.C. Rom\~ao, \zp C60 1993 757.

\bibitem{primo} D.~Bardin, A.~Leike and T.~Riemann, \pl B344 1995 383.

\bibitem{secondo} D.~Bardin, A.~Leike and T.~Riemann, in Proc. of
the `{\it Zeuthen Workshop on Elementary Particle Theory - Physics at
LEP~200 and Beyond}', ed. by T.~Riemann and J.~Bl\"umlein, Teupitz,
Germany, April 1994, \np B37 1994 274~(Proc.~Suppl.).

\bibitem{dima} D.~Bardin, A.~Leike and T.~Riemann, \pl B353 1995 513.
    
\bibitem{numerical} E.~Boos, M.~Sachwitz, H.J.~Schreiber and S.~Shichanin,
                 {\it Z. Phys} {\bf C61} (1994) 675; 
                 {\it ibidem}  {\bf C64} (1994) 391;
                 {\it ibidem}  {\bf C67} (1995) 613;
                 {\it Int. J. Mod. Phys.} {\bf A10} (1995) 2067;\\
M.~Dubinin, V.~Edneral, Y.~Kurihara and Y.~Shimitzu, \pl B329 1994 379;\\
F.A.~Berends, R.~Kleiss and R.~Pittau, {\it Nucl. Phys.} {\bf B424} 
(1994) 308; {\it ibidem} {\bf B426}
(1994) 344;  {\it Comput. Phys. Commun.} {\bf 85} (1995) 437.

\bibitem{GP} G. Passarino, \preprint\ hep-ph/9602302, February 1996;\\
             G. Montagna, O. Nicrosini, G. Passarino and F. Piccinini,
\pl B348 1995 178. 

\bibitem{GHS} P.~Grosse-Wiesmann, D.~Haidt and H.J.~Schreiber, in
              Ref.~\cite{ee500}.

\bibitem{NLCmio} S. Moretti, \preprint\ DFTT 40/95, DTP/95/58, June 1995
                 (to be published in {\it Z. Phys.} {\bf C}).

\bibitem{split1} A.~Ballestrero, E.~Maina and S.~Moretti,
                 {\it Phys. Lett.} {\bf B335} (1994) 460.

\bibitem{ISR} T.~Barklow, P.~Chen and W.~Kozanecki,
in Ref.~\cite{ee500}.

\bibitem{structure} F.A. Berends, W.L. van Neerven and G.J. Burgers,
{\it Nucl. Phys.} {\bf B297} (1988) 429; Erratum, {\it ibidem}
{\bf B304} (1988) 95;\\
E.A. Kuraev and V.S. Fadin, {\it Sov. J. Nucl. Phys.} {\bf 41}
(1985) 466;\\
G. Altarelli and G. Martinelli, Proceedings of the Workshop
`{\it Physics at LEP}', eds. J. Ellis
and R. Peccei, G\`eneva, 1986, CERN 86-02;\\
R. Kleiss, \np B347 1990 29.
 
\bibitem{Nicro} O.~Nicrosini and L.~Trentadue, {\it Phys. Lett.} {\bf B196}
(1987) 551; {\it Z. Phys.} {\bf C39}  (1988) 479.

\bibitem{count1} J. Fleischer, F. Jegerlehner, K. Kolodziej and G. J. van
Oldenborgh, {\it Comp. Phys. Comm.} {\bf 85} 
              (1994) 29.

\bibitem{count2} G. J. van Oldenborgh, 
\preprint\ INLO-PUB-95/04, March 1995, revised October 1995.

\bibitem{classification} 
F.A.~Berends, R.~Kleiss and R.~Pittau, {\it Nucl. Phys.} {\bf B424} 
(1994) 308;\\
D. Bardin, M. Bilenky, D. Lehner, A. Olchevski and T. Riemann,
in Proceedings of the `{\it Zeuthen Workshop on Elementary Particle Theory:
Physics al LEP~200 and Beyond}', eds. T. Riemann and J. Bl\"umlein,
{\it Nucl. Phys.} {\bf 37B} (1994) 148 (Proc. Suppl.).

\bibitem{ISRcomplete} D. Bardin, D. Lehner and T. Riemann, \preprint\
DESY 96--028, February 1996.

\bibitem{FSR} W. Beenakker, K. Kolodziej and T. Sack, \pl B258 1991 469;\\
       W. Beenakker, F.A. Berends and T. Sack, \np B367 1991 287;\\
       J. Fleischer, F. Jegerlehner and K. Kolodziej, \pr D476 1993 830.

\bibitem{benna} W. Beenakker, in Proceedings of the `{\it Zeuthen 
Workshop on Elementary Particle Theory:
Physics al LEP~200 and Beyond}', eds. T. Riemann and J. Bl\"umlein,
{\it Nucl. Phys.} {\bf 37B} (1994) 59 (Proc. Suppl.).

\bibitem{interf} V.S. Fadin, V.A. Khoze and A.D. Martin, {\it Phys.
Lett.} {\bf B311} (1993) 311; {\it ibidem} {\bf B320} (1994) 141;
\pr D49 1994 2247.

\bibitem{genuine} M. B\"ohm, A. Denner, T. Sack, W. Beenakker, F.A.
Berends and H. Kuijf, \np B304 1988 463;\\
J. Fleisher F. Jegerlehner and M. Zralek, \zp C42 1989 409.

\bibitem{tim} T.~Stelzer and W.F.~Long, {\it Comp. Phys. Comm.} {\bf 81} 
              (1994) 357.

\bibitem{HELAS} H.~Murayama, I.~Watanabe and K.~Hagiwara, HELAS: HELicity
                Amplitude Subroutines for Feynman Diagram Evaluations,
                {\it KEK Report} 91-11, January 1992.

\bibitem{VEGAS} G.P.~Lepage, {\it Jour. Comp. Phys.} {\bf 27} (1978) 192.

\bibitem{split2} S.~Moretti, {\it preprint} DFTT 69/94,
                DTP/95/02, December 1994
               (to be published in {\it Phys. Rev.} {\bf D}).

\bibitem{running} E.~Braaten and J.P.~Leveille, \pr D22 1980 715;\\
                  N.~Sakai, \pr D22 1980 2220;\\
                  T.~Inami and T.~Kubota, \np B179 1981 171;\\
                  M.~Drees and K.~Hikasa, \pl B240 1990 455;\\
                  S.G.~Gorishny, A.L.~Kataev, S.A.~Larin and
   	          L.R.~Surguladze, \mpl A5 1990 2703;\\
                  L.R.~Surguladze, \pl B341 1994 60.

\bibitem{cos} Z.~Kunszt and W.J.~Stirling, \pl B242 1990 507;\\
              N.~Brown, \zp C49 1991 657;\\
              V.~Barger and K.~Whisnant, \pr D43 1991 1443.

\bibitem{CDFtop} CDF~Collaboration,
{\it Phys. Rev. Lett.} {\bf 74} (1995) 2626.

\bibitem{D0top} D0 Collaboration, {\it Phys. Rev. Lett.} {\bf 74} (1995) 2632.

\bibitem{tt}  A.~Ballestrero, E.~Maina and S.~Moretti,
                 {\it Phys. Lett.} {\bf B333} (1994) 434.

\bibitem{vak} See for example:\\
              W.~Bernreuther {\it et al.}, in Ref.~\cite{ee500}, and
              References therein.

\bibitem{Bagliesi} G.~Bagliesi {\it et al.}, in Ref.~\cite{ee500}.
 
\bibitem{completo} See for example:\\
              W.~Beenakker {\it et al.}, {\it Report of the Working Group on
              WW Cross Sections and Distributions}, 
              in Ref.~\cite{lep2w}, and References therein;\\
              D.~Bardin {\it et al.}, {\it Report of the Working Group on
              Event Generators for WW Physics}, 
              in Ref.~\cite{lep2w}, and References therein;\\
              F.~Boudjema {\it et al.}, {\it Report of the Working Group on
              Standard Model Processes}, 
              in Ref.~\cite{lep2w}, and References therein;\\
              M.L. Mangano {\it et al.}, {\it Report of the Working Group on
              Event Generators for Discovery Physics}, 
              in Ref.~\cite{lep2w}, and References therein.

\bibitem{KLN} T. Kinoshita, {\it J. Math. Phys.} {\bf 3} (1962) 650;\\
              T.D. Lee and M. Nauenberg, {\it Phys. Rev.} {\bf 133B} (1964)
1549.

\end{thebibliography}
\end{document}